\documentclass[%
 reprint,
superscriptaddress,
 amsmath,amssymb,
 aps,onecolumn,prfluids
]{revtex4-2}

\usepackage{graphicx}% Include figure files
\usepackage{dcolumn}% Align table columns on decimal point
\usepackage{bm}% bold math
\usepackage{gensymb}
\usepackage{xcolor}

\begin{document}

% \preprint{APS/123-QED}

%%%%%%%%%%%
\title{Disk impact on a boiling liquid: Dynamics of the entrapped vapor pocket}

\author{Yee Li (Ellis) Fan}
\email{Contact author: ellisfan179@gmail.com}
\affiliation{% 
Physics of Fluids Group and Max Planck Center Twente for Complex Fluid Dynamics,
MESA+ Institute and J. M. Burgers Centre for Fluid Dynamics, University of Twente,
P.O. Box 217, 7500AE Enschede, The Netherlands}%
 
\author{Bernardo Palacios Muñiz}%
\affiliation{% 
Physics of Fluids Group and Max Planck Center Twente for Complex Fluid Dynamics,
MESA+ Institute and J. M. Burgers Centre for Fluid Dynamics, University of Twente,
P.O. Box 217, 7500AE Enschede, The Netherlands}%

\author{Nayoung Kim}%
\affiliation{% 
Physics of Fluids Group and Max Planck Center Twente for Complex Fluid Dynamics,
MESA+ Institute and J. M. Burgers Centre for Fluid Dynamics, University of Twente,
P.O. Box 217, 7500AE Enschede, The Netherlands}%

\author{Devaraj van der Meer}
\email{Contact author: d.vandermeer@utwente.nl}
\affiliation{% 
Physics of Fluids Group and Max Planck Center Twente for Complex Fluid Dynamics,
MESA+ Institute and J. M. Burgers Centre for Fluid Dynamics, University of Twente,
P.O. Box 217, 7500AE Enschede, The Netherlands}%

\date{\today}% It is always \today, today, but any date may be explicitly specified
%%%%%%%%%

\begin{abstract}

Upon the impact of a flat disk on a boiling liquid, i.e., a liquid that is in thermal equilibrium with its own vapor, a thin vapor layer is entrapped under the disk. Due to the tendency of vapor to undergo phase change under pressure variation upon impact, the dynamics of this entrapped vapor pocket are different from those of a non-condensable air pocket. In this work, we experimentally investigate the dynamics of the entrapped vapor pocket, more specifically its time evolution and its subsequent influence on the hydrodynamic loads at different equilibrium ambient temperatures and impact velocities. We find that the retraction of the vapor pocket at high ambient temperature and small impact velocity is slow, occurring from the disk edge, and driven by the dynamic pressure $\rho_{\text{L}}U_0^2$. In contrast, at lower ambient temperatures and large impact velocities, after a short initial stage, the vapor pocket will collapse rapidly due to condensation. This scenario is confirmed by conducting experiments where, by heating the disk, the vapor pocket collapse is observed to slow down. We attribute this to the vaporization of liquid near the three-phase contact line region that frustrates the condensation process and reduces the impact pressure on the disk. The violent collapse of the vapor pocket may impart additional instantaneous momentum, but the overall pressure and force impulses are still found to be closely associated with the liquid added mass. Finally, we found that at a high tilt angle of the disk, the three-phase contact line movement over the disk surface may hinder the proper entrapment and compression of the vapor pocket, which results in a lower central impact pressure as rapid condensation at the central disk region does not occur.

\end{abstract}
\maketitle

\section{\label{sec:Intro} Introduction}

The role of the interstitial medium during solid-liquid or liquid-liquid impact is often crucial as it influences the overall dynamics of the impact process. Here, the most investigated interstitial medium is air, a non-condensable gas. The air phase reduces the maximum dynamic loads during solid-to-liquid impact \cite{verhagen1967impact,kim2021water}, degrades the printing quality during inkjet printing \cite{lohse2022fundamental} and produces the sound we hear from a dripping tap \cite{phillips2018sound}. In particular, the dynamics of the air layer entrapped during solid-liquid impact has been investigated extensively since the early experimental studies on the air cushioning effect \cite{chuang1966experiments,verhagen1967impact}. The volume of entrapped air, which is determined by the geometry of the impacting body, is of key interest as it directly influences the resulting hydrodynamic loads on the impacting body. A recent study showed that a slightly convex nose that entraps a thinner layer of air as compared to a flat-bottom body results in a higher peak impact load \cite{belden2024water}. The wetting of the liquid or the retraction of the air layer is also widely investigated, especially in the case of a wedge or a cone \cite{jain2022wedge,carrat2023air}, for their role on the pressure distribution on the impacting body resulting from Wagner's theory \cite{wagner1932stoss}. The viscosity and surface tension of the liquid are found to influence the retraction rate of the entrapped air upon impact \cite{thoroddsen2005air}. In general, entrapping air is effective in mitigating the maximum load on the impacting body because the loading duration will be prolonged. Due to the cushioning effect of the non-condensable air, for impact at moderate impact velocity $\mathcal{O}(1)$ m/s one does not necessarily need to account for the liquid compressibility. However, in the case of a boiling liquid, i.e., a liquid that is in thermal equilibrium with its own vapor, the interstitial medium consists of vapor instead of non-condensable air. Vapor, being condensible, may undergo a sudden phase change upon impact, frustrating the cushioning effect as the vapor disappears rapidly. Consequently, liquid compressibility will come into play and influence the impact dynamics.

While an air bubble existing in a liquid bulk usually simply deforms and ruptures into smaller bubbles upon compression \cite{bird2010daughter,jain2022controlled}, a vapor bubble within a liquid may implode \cite{benjamin1966collapse,prosperetti2017vapor}, where it rapidly contracts and depending on the geometry, may form a self-piercing jet upon compression. This so-called cavitation process and the subsequent implosion of the vapor bubble is known to induce shock waves \cite{supponen2017shock} which may cause different type and degree of local damage to surrounding objects, depending on the boundary conditions \cite{bokman2023high,sagar2024dynamics}. Impact events involving entrapped vapor are typically encountered in a cryogenic fuel transportation tank containing, e.g., liquid hydrogen (LH2) or liquid natural gas (LNG), where the liquefied gaseous fuel experiences sloshing during transportation \cite{dias2018slamming}. Since in such a system the liquid can be assumed to be in equilibrium with its own vapor (at some temperature and corresponding vapor pressure, where the latter is typically atmospheric), we name it a boiling liquid, for short. Due to its relevance to cryogenic fuel transportation, some recent studies have been carried out on a large wave-impact scale to investigate the influence of phase change on the sloshing dynamics \cite{hicks2018lng,lee2021experimental,ancellin2012influence}. Reported results are however sometimes rather contradictory and less conclusive on the effect of phase change on the impact loads. Sloshing occurs over a wide range of length scales \cite{hicks2018lng}; therefore, in order to understand the global dynamics, it will require a fundamental understanding of the boiling liquid impact process, starting from the small length scale, which is knowledge that is currently largely missing.

In this work, we study the impact of a horizontal circular flat disk onto a boiling liquid, where we focus on the evolution of the vapor pocket entrapped under the disk upon impact and how its behaviour influences the solid-liquid interaction, and subsequently results in different hydrodynamics loads.  We explore the effect of the ambient temperature of the system, the impact velocity, the temperature and the tilting angle of the disk through a series of experiments in a temperature-controlled sealed chamber.

The paper is organized as follows: In Section \ref{sec:setup}, we give an overview of the experimental setup and the measurement technique. Section \ref{sec:Entrapment} describes the pre-impact process that results in the entrapment of the vapor pocket. Subsequently, Section \ref{sec:Dynamics} focuses on the post-impact dynamics of the vapor pocket, including the sometimes rapid evolution of the vapor pocket during the early stage of impact and the resulting hydrodynamics loading exerted on the disk and in the liquid bulk. Section \ref{sec:Evaporation} explores the effect of vaporization on the vapor pocket by heating the disk and shows that vaporization of liquid at the contact line can counteract the rapid collapse of the vapor pocket. Finally, in Section \ref{sec:tilt}, we look into the influence of the tilting angle of the disk on the vapor pocket behaviour and the impact loadings, before proceeding to conclude our work and to provide some perspective on how understanding of boiling liquid impact can offer new insights into the century-old problem of solid-liquid impact in Section \ref{sec:Conclusion}.

\section{\label{sec:setup} Experimental set-up}

\begin{figure*}
\includegraphics[clip,trim=0cm 0cm 0cm 0cm, width=.6\textwidth]{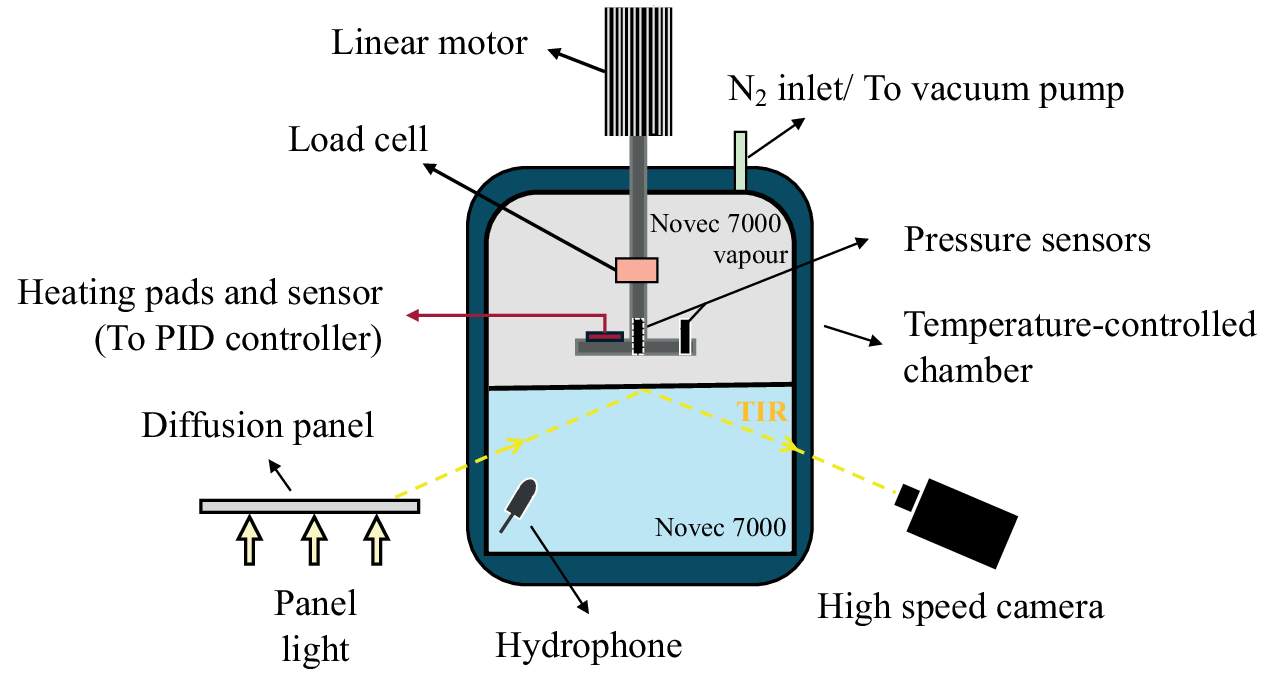}
\caption{\label{fig:setup} Schematic of the experimental setup.}
\end{figure*}

As depicted in Fig. \ref{fig:setup}, a boiling liquid system where the liquid is in thermal equilibrium with its own vapor is established in a 30 $\times$ 30 $\times$ 30 cm$^3$ vacuum-sealed chamber. The procedure to realize the boiling liquid system is as follows: The system is first evacuated with a vacuum pump and flushed with nitrogen gas twice before the working liquid Novec 7000 (C$_4$F$_7$OH$_3$) is released into the system, vaporizing and filling the chamber until a liquid height $H_\text{L} \approx$ 0.175 m is reached. A circular flat steel disk (RVS 316) of radius $R_0 = 0.040$ m and thickness $d =$ 0.0050 m is connected to a linear motor (Dunkermotoren, STB2506S) via a rod and impacted vertically onto the liquid bulk at different impact velocities $U_0$ ranging from 0.5 m/s to 2.0 m/s. The ambient temperature $T_{0}$ in the system is varied from $\approx 11.5^{\circ}$C to $\approx 24.5^{\circ}$C, which corresponds to a saturation pressure ranging from $p_{\text{v,0}}$ $\approx$ 420 to $\approx$ 740 mbar. Note that the saturation pressure reached in our system is always higher than the reference saturation pressure reported in \cite{widiatmo2001equations} by approximately 50 mbar, most likely due to the presence of dissolved air in the liquid.

To measure the hydrodynamic loads during impact, two Kistler pressure sensors (Type 601C) are flush-mounted at the disk center and near the disk edge to record the local impact pressure at an acquisition rate of 200 kHz. In addition, a load cell (FUTEK, LRF350) with a maximum tension/compression loading capacity of $\pm$ 890 N is used together with an analogue amplifier with current output (FUTEK, IAA2000) to measure the impact load exerted on the entire disk surface at 20 kHz. The load cell is installed along the rod that connects the disk with the linear motor. A hydrophone (RESON, TC4013) with a frequency range of 1 Hz to 170 kHz and a receiving sensitivity of $-211$ dB $\pm$ 3 dB re 1V$/\mu$Pa is placed about 0.04 m under the free liquid surface and 0.12 m from the center of the disk to measure the pressure changes in the liquid bulk. The pressure signal from the hydrophone is recorded at a 1 MHz sampling frequency by a DAQ device (National Instrument, USB-4432). For synchronization, the digital signals from the pressure sensors and the load cell are also sent to the DAQ device. A heating system comprising two 31 mm $\times$ 19.1 mm polyimide heating pads (Thermo Tech, 14.4W) and a PT100 sensor, controlled by a PID control system, is used to regulate the temperature of the impacting disk. The self-adhesive heating pads and the PT100 sensor are attached to the back of the disk. With good thermal conductivity of steel and sufficient heating time between experiments, the temperature on the impacting surface is assumed to be the same as that measured at the back of the disk.

The total internal reflection (TIR) technique described in \cite{jain2021total} is utilized to quantify the surface deformation before the impact and to visualize the dynamics of the entrapped vapor pocket under the disk upon impact. The impact process is viewed at an angle of $\approx 23 ^{\circ}$ with the free liquid surface, using a high-speed camera (Photron Nova 16) imaging at 30k, 40k or 48k fps and a light illuminating from the opposite side of the camera. A reference grid pattern is placed over the diffusion panel above the light, and the pattern is reflected by the free liquid surface towards the camera. As the disk approaches the free liquid surface, the deformation of the liquid surface causes the reflected pattern to distort. Based on this distortion, the deformation of the liquid surface prior to impact can be quantified using the method formulated in \cite{jain2021total}. When the disk is in contact with the liquid surface, the reflection of the liquid surface is obstructed and the wetted area appears black in the recorded images. 

\section{\label{sec:Entrapment} Entrapment of a vapor pocket under the disk}

Before exploring the dynamics of the entrapped vapor pocket upon impact, we first determine and explain how the vapor pocket is actually formed. Due to the deformation of the free liquid surface prior to impact, a thin layer of vapor, known as the vapor pocket hereafter, is entrapped under the impacting disk upon impact. Similar to the air layer trapped when a liquid droplet impacts onto a solid substrate or into a liquid pool, as the vapor between the disk and the liquid surface is being squeezed out by the descending disk, local pressure build-up causes the liquid surface to depress \cite{bouwhuis2015initial}. In the meantime, the liquid surface around the disk edge is elevated due to rapid exiting gas flow which induces the growth of Kelvin-Helmholtz instability under the disk edge \cite{jain2021KH,fan2024air}. Therefore, the initial contact between the disk and the liquid surface occurs near the disk perimeter upon impact, resulting in the entrapment of a vapor pocket. 

\subsection{\label{sec:deformation} Free surface deformation prior to impact}

\begin{figure*}
\includegraphics[clip, width=1\textwidth]{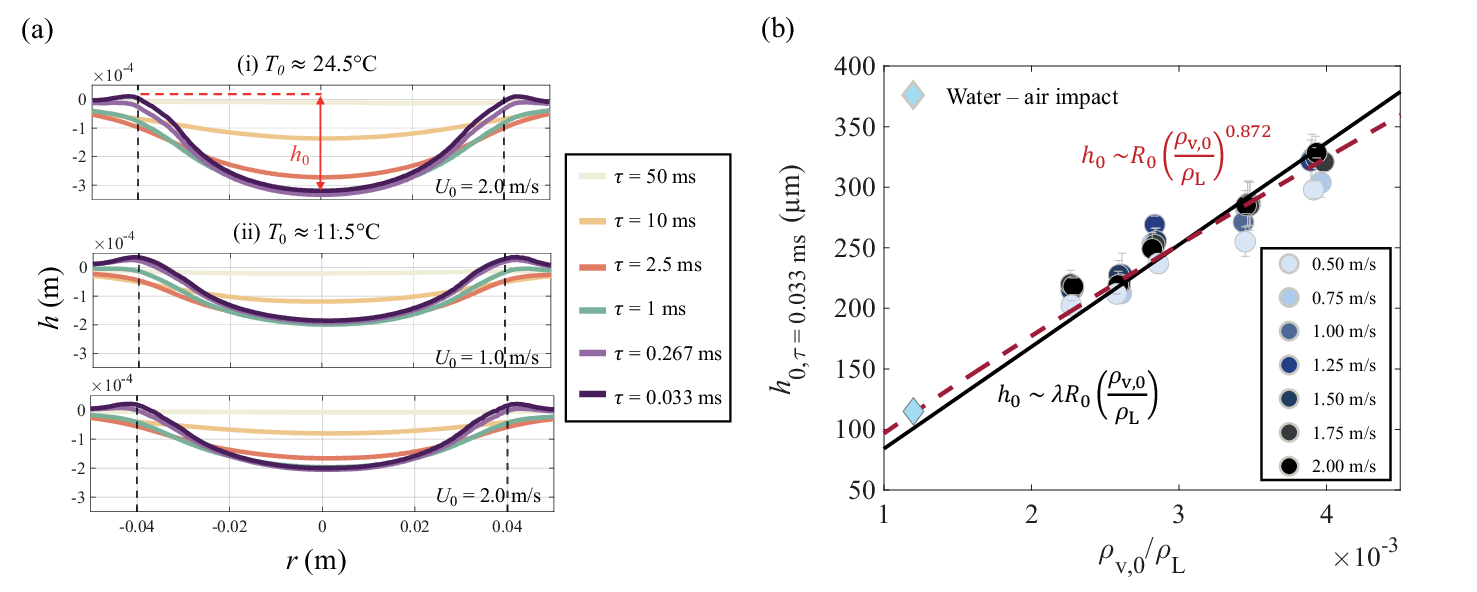}
\caption{\label{fig:h0} (a) Cross section of the temporal deformation of the free liquid surface before the impact of a $R_0 =$ 0.040 m circular flat disk at (i) $T_0 \approx 24.5^{\circ}\text{C}$ and $U_0 =$ 2.0 m/s and (ii) $T_0 \approx 11.5^{\circ}\text{C}$ and $U_0 =$ 1.0 and 2.0 m/s. The vertical dashed lines denote the location of the disk edge and $\tau$ is defined as the time before impact, where $\tau$ = 0 is the moment at which the disk makes initial contact with the deformed surface around the disk edge. The initial thickness of the entrapped vapor pocket $h_0$ is taken as the difference between  $h(r = R_0$) and $h(r = 0)$ at $\tau =$ 0.033 ms. (b) Plot of the initial thickness of the entrapped vapor pocket $h_0$ against the density ratio between the vapor and liquid $\rho_{\text{v,0}}/\rho_{\text{L}}$. The black solid line shows the direct proportionality of $h_0$ with {$R_0 \left(\rho_{\text{v,0}}/\rho_{\text{L}}\right)$, where $\lambda = 2.1$} is a fitted prefactor based on the experimental data. Meanwhile, the red dashed line $h_0 \sim R_0 (\rho_{\text{v,0}}/\rho_{\text{L}})^{0.872}$ represents the dependence of $h_0$ on the density ratio at $t^{\dagger} =$ 0.9998 predicted by the integration of Eq. (5.6) in \cite{peters2013splash}. The light blue {diamond} marker at $\rho_\text{v,0}/\rho_\text{L} = 1.2 \times 10^3$ is the initial thickness of an entrapped air pocket under a $R_0 = 0.040$ m flat disk before impacting towards a water bath at $U_0 = 1.0$ m/s, taken from \cite{jain2021total}.
}
\end{figure*}

With the TIR technique discussed in Section \ref{sec:setup} and detailed in \cite{jain2021total}, the free surface before impact is reconstructed. The cross-section of the temporal evolution of the free liquid surface at two selected ambient temperatures ($T_0 \approx$ 11.5$^{\circ}\text{C}$ and 24.5$^{\circ}\text{C}$) and impact velocities ($U_0 =$ 1.0 and 2.0 m/s) are shown in Fig. \ref{fig:h0}a. Here, $\tau$ is the amount of time remaining before impact, where $\tau$ = 0 is the moment when the disk makes initial contact with the deformed surface around the disk edge. As seen in Fig. \ref{fig:h0}a, the depression of the liquid surface under the disk center ($r$ = 0 m) first increases with time and saturates at around $\tau =$ 1 ms for all three cases, consistent with what is observed in air \cite{peters2013splash,jain2021air}. However, the saturation depth of the depression at $T_0 \approx 24.5^{\circ}\text{C}$ is observed to be larger as compared to that at $T_0 \approx 11.5^{\circ}\text{C}$. At a similar time as that when the saturation depth is reached, around the disk edge an instability is seen to start growing significantly which is connected to the increase of the exit gas velocity with the descent of the disk \cite{jain2021KH,van2022linear}. Comparing Figs. \ref{fig:h0}a(i) and (ii), the saturation depth appears to vary with the change in ambient temperature, but is less influenced by the change in impact velocity (Fig. \ref{fig:h0}a(ii)). The strong dependence of the saturation depth on the ambient temperature is further highlighted in Fig. \ref{fig:h0}(b), which will be discussed in further detail below.

Summarizing the above discussion, the initial thickness of the entrapped vapor pocket $h_0$ can be estimated from the maximum deformation of the liquid surface at $\tau =$ 0.033 ms, one frame before impact. It is taken as the difference between the maximum height of the elevated surface at the disk edge $h(r = R_0$) and the minimum depth of the depressed liquid surface $h(r = 0)$, as indicated by the red arrow in Fig. \ref{fig:h0}(a)(i). Fig. \ref{fig:h0}b shows the initial thickness of entrapped vapor pocket $h_0$ (i.e., the maximum surface deformation) plotted against the density ratio between the vapor and liquid $\rho_{\text{v,0}}/\rho_{\text{L}}$ for different ambient temperatures, where the latter is computed based on the thermal properties of Novec 7000 provided in \cite{aminian2022ideal,ohta2001liquid}. The density ratio between the vapor and liquid increases with the ambient temperature due to higher ambient saturated vapor pressure, which corresponds to a higher vapor density. The liquid density also decreases slightly with increasing ambient temperature, but this is a minor effect. It can be observed that the thickness of the vapor pocket increases from $\sim$ 200 $\mu$m to $\sim$ 320 $\mu$m with an increasing density ratio between the vapor and the liquid from 2.3 $\times 10^{-3}$  to 3.9 $\times 10^{-3}$. However, the thickness of the vapor pocket $h_0$ varies only slightly (with $< 35~\mu$m deviation) at a fixed density ratio, despite the impact velocity being changed fourfold from 0.5 m/s to 2.0 m/s. 

In potential flow, dimensional analysis provides that at a fixed distance $H$ of the disk to the undisturbed fluid, the liquid interface deformation is only a function of the density ratio, i.e., ${h_0}/{R_0} = f({H}/{R_0},{\rho_\text{v,0}}/{\rho_\text{L}})$. Assuming that the maximum deformation is found for the same $H/R_0$, one would therefore expect that ${h_0}/{R_0} = f({\rho_\text{v,0}}/{\rho_\text{L}})$. Since the impulse exerted on the liquid interface is proportional to $\rho_\text{v}$, one is tempted to try a proportionality, $h_0/R_0= \lambda \rho_{\text{v,0}}/\rho_{\text{L}}$. The initial thickness of the vapor pocket $h_0$ follows this proportional dependence on the density ratio reasonably well with a best-fitted prefactor {$\lambda = 2.1$} as shown as the black solid line in Fig. \ref{fig:h0}b. A more advanced analysis is found in \cite{peters2013splash}, where again under the potential flow assumption, the deformation of the liquid surface under an impacting circular flat disk is studied, and a differential equation for the vertical deformation of the liquid surface is obtained. The time evolution of the vertical displacement of the liquid surface $h$ can be obtained by integrating Eq. (5.6) in \cite{peters2013splash} numerically, with the non-dimensional time after impact $t^{\dagger}$ defined in \cite{peters2013splash} corresponding to $t^{\dagger} = 1 - \tau U_0/R_0$.

By taking $h$ at $t^{\dagger} =$ 0.9998, the integration results in $h_0 \sim R_0 (\rho_{\text{v,0}}/\rho_{\text{L}})^{0.872}$, which is plotted as the red dashed line in Fig \ref{fig:h0}b, agreeing well with current experimental data. Good agreement between experimental results and theoretical analysis based on potential flow suggests that the deformation of the free liquid surface during boiling liquid impact is mainly an inertial effect, where vapor viscosity and capillary effects are less influential, at least for the parameter range studied in this work. Phase change does not appear to play a role in the pre-impact dynamics. Note that, although overall there is little variation in $h_0$, at small impact velocities, especially at $U_0 =$ 0.5 m/s, the initial thickness of the vapor pocket $h_0$ tends to be slightly thinner for all density ratios, which could be due to the effect of surface tension trying to restore the deformed surface for these low impact velocities \cite{bouwhuis2012maximal,bouwhuis2015initial}. Finally, we have added a data point for the thickness $h_0$ measured for an equally sized disk impacting on water in air, at the air-water density ratio $\rho_\text{v,0}/\rho_\text{L} = 1.2 \times 10^{-3}$ taken from \cite{jain2021total}, which matches with the scaling law found for Novec 7000.

\section{\label{sec:Dynamics} Post-impact dynamics of the entrapped vapor pocket}

\subsection{\label{sec:Observation} General experimental observations}

\begin{figure*}
\includegraphics[clip,trim=0cm 0cm 0cm 0cm, width=.8\textwidth]{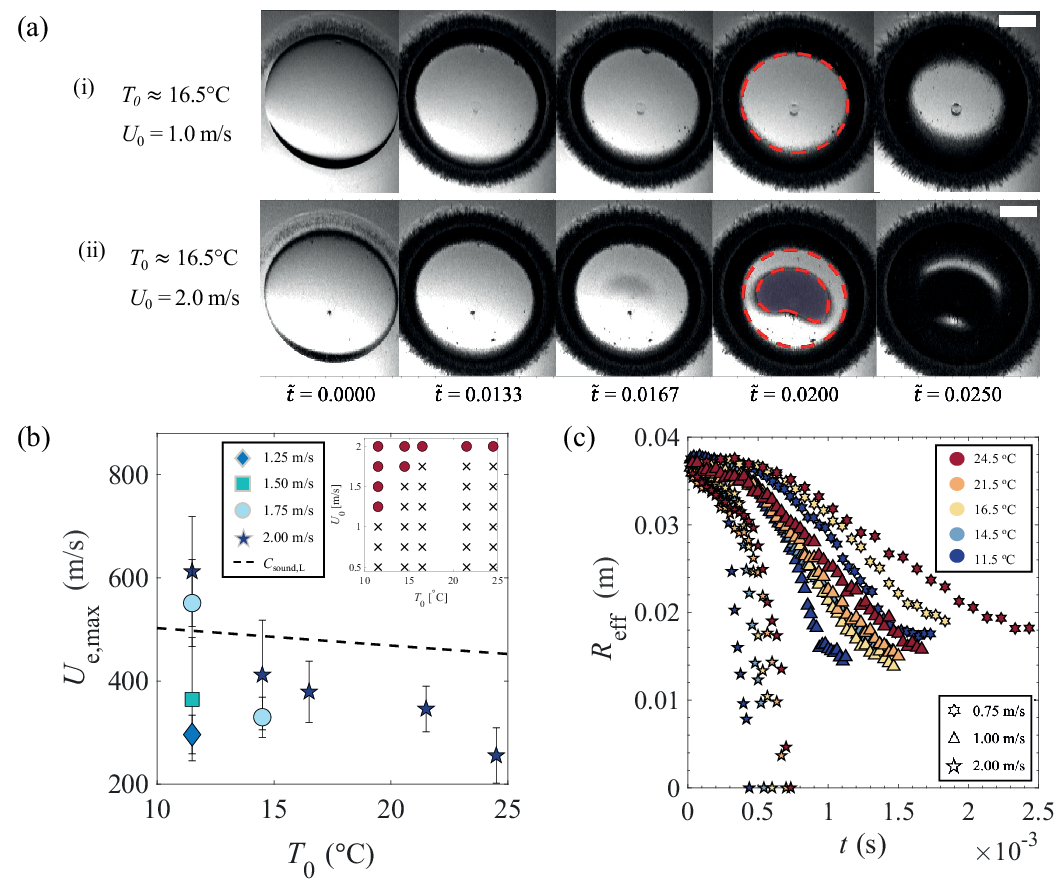}
\caption{\label{fig:Reff} (a) Six (re-aspected) snapshots of the evolution of the entrapped vapor pocket at (i) $U_0$ = 1.0 m/s and (ii) $U_0$ = 2.0 m/s at $T_0 \approx$ 16.5$^{\circ}\text{C}$. The time $\tilde{t}$ is the time after impact $t$ normalized by the inertial time scale $t_{\text{i}} = R_0/U_0$, where $\tilde{t}$ = 0 is the moment at which the disk edge first contacted the deformed liquid surface. The scale bar indicates a length of 20 mm. (b) The maximum expansion speed $U_{\text{e,max}}$ of the wetted area from the central disk region for selected $T_{0}$ and $U_0$ where the entrapped vapor pocket collapses. The inset shows a phase diagram where red circular markers indicate that the vapor pocket is observed to collapse at the central region in the experiment for the corresponding ambient temperature $T_0$ and impact velocity $U_0$, while the cross markers indicate cases where no collapse at the central region was registered. The black dashed line indicates the speed of sound $C_{\text{sound,L}}$ in liquid Novec 7000 at different temperatures from \cite{aminian2022ideal}. (c) Plot of the effective radius $R_{\text{eff}}$ of the vapor pocket against time after impact $t$. Here, symbols indicate the different impact velocities and colors the different ambient temperatures (see legend).} 
\end{figure*}

Now that we have studied the formation of the vapor cavity prior to impact, we turn to its dynamics just after the impact. Referring to Fig \ref{fig:Reff}a, where the impacting disk first makes contact with the deformed liquid surface at $\tilde{t} = tU_0/R_0 =$ 0 (equivalent to $\tau = 0$), the disk touches the liquid surface from one edge (from the top in these example cases) due to a small but unavoidable tilt of the disk. For the results discussed in this Section, the tilting angle of the impacting disk $\alpha$ is almost constant at 0.25 $\pm$ 0.05$^{\circ}$. This degree of tilting has only a minor influence on the collapse behaviour of the vapor pocket and it does not transform into the type of impact described in \cite{speirs2021cavitation}, where the three-phase contact line moves across the impacting surface. The effect of tilting will be discussed in more detail in Section \ref{sec:tilt}.

As the disk further descends, all the edges are wetted evenly and the three-phase contact line, indicated by the circular red dashed lines, retracts inward toward the disk center. This accurately describes the series of events for $U_0 = 1.0$ m/s in Fig. \ref{fig:Reff}a(i), but for the higher impact velocity $U_0 = 2.0$ m/s of Fig \ref{fig:Reff}a(ii), there is an additional phenomenon, where the vapor pocket is seen to collapse, initiating solid-liquid contact at the central region of the disk at $\tilde{t} =$ 0.0167 and subsequently expanding rapidly outward towards the disk edge, as highlighted by the shaded area at $\tilde{t} =$ 0.02. We found that such collapse behaviour of the vapor pocket that is observed at high impact velocities and low ambient temperature is induced by the condensation of the vapor pocket \cite{li2025impactpressureenhancementdisk}. A greater compression at high impact velocity together with a smaller vapor density and a smaller initial thickness $h_0$ at low ambient temperature cause the vapor to condense more quickly, accelerating the contraction of the vapor pocket. The dynamics of the entrapped vapor pocket upon impact can be quantified by the expansion of the wetted area at the central disk region and the retraction of the three-phase contact line as discussed in the following sections.

\subsection{\label{sec:Expansion} Expansion speed of the wetted area}

The collapse of the vapor pocket from the central disk region during boiling impact at low ambient temperature and high impact velocity is quantified by tracing the evolution of the wetted area $A_\text{w}$ (shaded region in  Fig. \ref{fig:Reff}a(ii), $\tilde{t} =$ 0.02) with time. By approximating the wetted area as circular, an effective radius is defined as $R_\text{w} = \sqrt{A_\text{w}/\pi}$ from which the expansion speed is approximated as $U_{e} = \Delta R_\text{w}/\Delta t$, where $\Delta R_\text{w} = R_\text{w,i+1} - R_\text{w,i}$ and $\Delta t = t_{\text{i+1}} - t_{\text{i}}$, with $i$ being the frame number. Fig. \ref{fig:Reff}b shows the maximum expansion speed $U_{\text{e,max}}$ plotted against the ambient temperature $T_{0}$ for selected high-impact velocities where the vapor pocket collapses (see inset for the cases that collapse or do not). The maximum expansion speed is usually measured within the first three frames ($\leq$ 0.1 ms) after the vapor pocket starts to collapse. At $U_0 =$ 2.0 m/s, when the ambient temperature decreases, the maximum expansion velocity is seen to increase. Keeping the temperature constant at the lowest attainable value $T_{0} \approx 11.5^{\circ}$C, the maximum expansion velocity decreases when the impact velocity decreases. At high impact velocity and low ambient temperature, the expansion speed can exceed the speed of sound in liquid Novec 7000 $C_{\text{sound,L}}$, where the latter is shown as the black dashed line in the plot. 

\subsection{\label{sec:retraction} Overall retraction of vapor pocket}

Similar to the previous subsection, the overall retraction of the vapor pocket can be quantified by tracing the movement of the three-phase contact line of the vapor pocket. For a non-collapsing vapor pocket as in Fig. \ref{fig:Reff}a(i), the effective radius $R_\text{eff}$ is essentially the temporal radius of the circle indicated by the red dashed line. When the vapor pocket collapses from the central disk region as in the case of Fig. \ref{fig:Reff}a(ii), the effective radius of the remaining vapor pocket is estimated by deducting the wetted area $A_\text{w}$ from the area $A$  bounded by the circular red dashed line, i.e., $A - A_\text{w} = \pi R_{\text{eff}}^2$. In Fig. \ref{fig:Reff}c, we show the effective radius of the vapor pocket as a function of time for several representative ambient temperatures (colors) and impact velocities (symbols). For the non-collapsing cases, we stop tracing when the radius starts to remain more or less constant. The plot for all ambient temperatures and impact velocities can be found in the Supplementary Material \cite{supplemental}. It was observed that, for the same impact velocity at different ambient temperatures, the initial retraction stage begins at a similar rate but deviates at a later stage, especially for the lowest ambient temperature ($T_{0} \approx 11.5^{\circ}$C). As expected, rapid retraction occurs at high impact velocity due to the collapse of the vapor pocket from the central disk region. 

\begin{figure*}
\includegraphics[clip,trim=0cm 0cm 0cm 0cm, width=.6\textwidth]{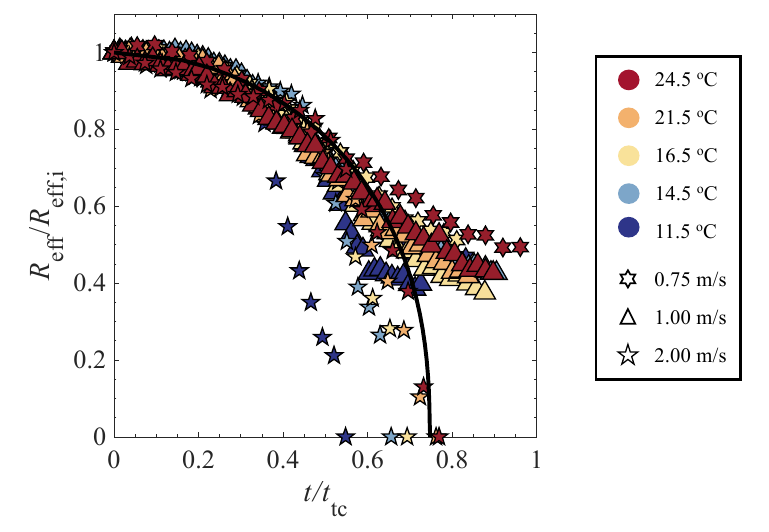}
\caption{\label{fig:normR} Normalized radius {$R_\text{eff}/R_\text{eff,i}$} of the entrapped vapor pocket against normalized time {$t/t_\text{tc}$} for selected representative ambient temperatures and impact velocities from Fig. \ref{fig:Reff}c. The radius of the vapor pocket $R_{\text{eff}}$ is normalized by its initial radius $R_{\text{eff,i}}$ while the time after impact $t$ is normalized by the Rayleigh collapse time $t_{\text{tc}}$ (see text). The black solid line is the solution of the Eq. \eqref{eq:rayleigh}. The data for different ambient temperatures and impact velocity collapse well at the initial stage of the retraction.}
\end{figure*}

Typically, for disk impact on a liquid at low impact velocity ($U_0 < 0.2$ m/s) \cite{jain2021air,carrat2023air} or for liquid droplet impact on a solid substrate \cite{thoroddsen2005air,kim2021water}, the retraction of the entrapped air between the solid and liquid is described as a balance between liquid inertia and surface tension, which is a Taylor-Culick type retraction. However, in our experiment, due to the low surface tension of Novec 7000 (0.0124 N/m) and comparatively high impact velocity ($U_0 \geq 0.5$ m/s), the retraction of the vapor pocket is expected to be driven by the pressure, similar to a Rayleigh collapse. Approximating the entrapped vapor pocket as a spherical cap with a base radius $R_0$ and height $h_0$, the volume of the spherical cap is $V_{\text{cap}} = \frac{1}{6} \pi h_0 (3R_0^2 + h_0^2)$. With this, we compute the equivalent radius $R_{\text{equiv}}$ of the vapor pocket assuming its volume to be spherical as
\begin{equation}
    R_{\text{equiv}} = \sqrt[3] {\frac{1}{8} h_0 \left(3R_0^2 + h_0^2 \right)}
\end{equation}
Assuming the internal pressure of the sphere (and thus the vapor pocket) to be zero, the time of collapse $t_{\text{collapse}}$ to a given fraction of its original radius $R$ as derived in \cite{rayleigh1917viii} is:
\begin{equation}\label{eq:rayleigh} 
    t_{\text{collapse}} = R_0 \sqrt{\frac{3\rho_{\text{L}}}{2P}} \int^{1}_{\tilde{\beta}} \frac{\tilde{\beta}^{3/2}}{(1 - \tilde{\beta}^3)^{1/2}} d\tilde{\beta}
\end{equation}
where $R_0$ is the initial radius of the sphere, $\tilde{\beta} = R/R_0$ and $P$ is typically the pressure at infinity, which can be taken as the driving pressure in our case. The time required for total collapse of the vapor pocket $t_{\text{tc}}$ from $R_0 = R_{\text{equiv}}$ to $R = 0$ can be obtained as:
\begin{equation}
    t_{\text{tc}} = 0.915 R_{\text{equiv}} \sqrt{\frac{\rho_{\text{L}}}{P}}
\end{equation}
Considering the driving pressure from the impact, we take $P \approx \rho_{\text{L}}U_0^2$ and normalize the plot of $R_{\text{eff}}$ against $t$ from Fig. \ref{fig:Reff}c with the initial radius of the vapor pocket $R_{\text{eff,i}}$ and $t_{\text{tc}}$ respectively. The result is found in Fig. \ref{fig:normR}, where, again we only present a subset of the data, but the normalized plot for all $T_{0}$ and $U_0$ covered in the experiment can be found in the Supplementary Material \cite{supplemental}. The full solution of Eq. \eqref{eq:rayleigh} can be obtained by numerical integration and is plotted in Fig. \ref{fig:normR} for comparison with the data. As seen from Fig. \ref{fig:normR}, the curves for the different ambient temperatures and impact velocities collapse reasonably well, especially during the initial retraction stage, where the data is also reasonably well described by Eq. \eqref{eq:rayleigh}. This indicates that the retraction at the initial stage is driven by the dynamic pressure and the pressure inside the vapor pocket is comparatively small. The impact velocity influences the degree to which the vapor pocket is being compressed, whereas the main effect of the ambient temperature is in determining the initial size of the vapor pocket which results in different $R_{\text{equiv}}$. At moderate impact velocities ($U_0 = 0.75$ m/s and 1.0 m/s), the retraction is observed to be slowed down at a later time. This is likely due to the internal pressure in the vapor pocket, which is ignored in Rayleigh collapse but is expected to increase adiabatically (from its initial value $p_{\text{v,0}}$) as the pocket is compressed, as long as condensation can be neglected. This internal pressure build-up will then push back, resisting the compression induced by the external pressure from the liquid phase. Thus, in later stages, the time evolution of the vapor pocket is slower than predicted by Eq. \eqref{eq:rayleigh} at these moderate and small speeds.

In contrast, at the highest impact velocity $U_0 = 2.0$ m/s, the collapse of the vapor pocket happens at a larger rate than that of a Rayleigh collapse. The fast condensation of the vapor pocket causes it to collapse from the center and to eventually break down into smaller bubbles or bubble fragments. It is remarkable that this collapse is faster than a Rayleigh collapse, which may be related to the fact that the pocket is in fact thin and may be squeezed as soon as its vapor contents have disappeared through condensation. Most importantly, those cases where a strong condensation-induced vapor pocket collapse from the central region was observed, stand out by first following a similar time evolution as the other cases but then at $t/t_\text{tc} \approx 0.4 - 0.5$, abruptly decaying to zero. An interesting intermediate case is that observed for $T_0 \approx 11.5^{\circ}$C, $U_0 = 1.0$ m/s which first appears to plummet like some of the $U_0 = 2.0$ m/s cases, but then recovers and tends away from fast decay, following a similar path as that followed for the same impact speed at higher ambient temperatures. This may indicate vapor condensating at first until a point has been reached where the latent heat of condensation cannot be transported into the liquid any longer, after which the contents of the pocket start to behave more like a non-condensable gas would do.

\subsection{\label{sec:Load} Resulting hydrodynamics loads}

Now that we have obtained insight in the dynamics of the vapor pocket, we turn to a more extensive study of the pressures and forces occurring during impact, both those observed on the disk and the resulting pressures in the bulk of the liquid, as recorded by the hydrophone. Then, finally, we will turn to time-integrated measurements, most importantly, the pressure impulse.

\subsubsection{Impact pressure and pressure in the liquid bulk}
\begin{figure*}
\includegraphics[clip, width=.9\textwidth]{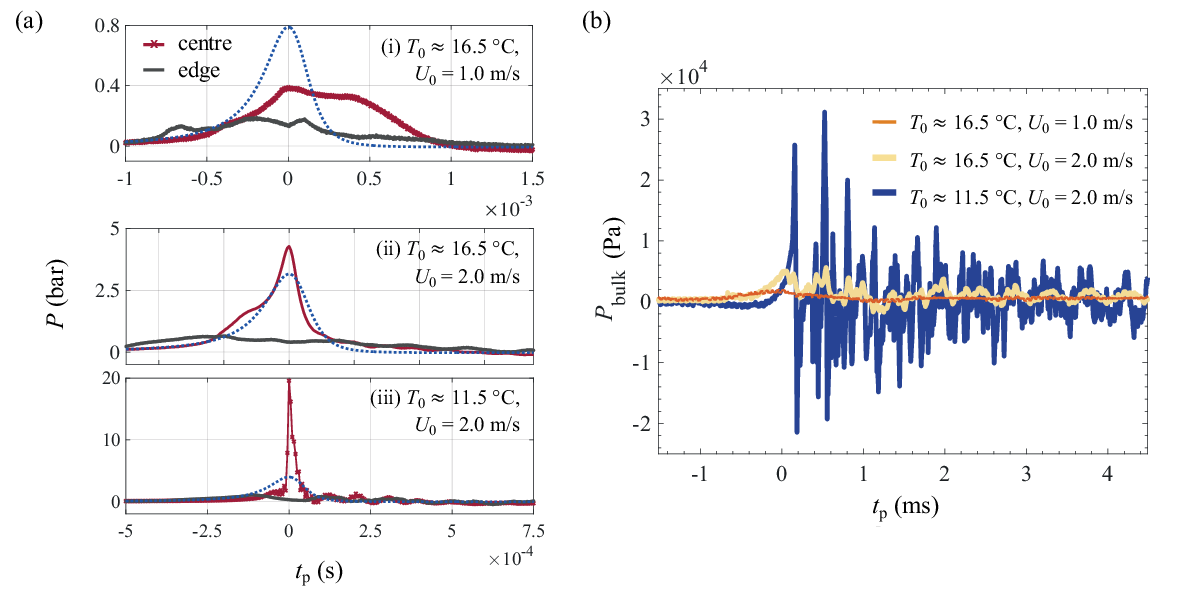}
\caption{\label{fig:Pressure} (a) Time series of the pressure signal $p(t)$ at the center (red lines) and near the edge (black lines) of the disk at $T_{0} \approx$ 16.5 $^{\circ}$C and (i) $U_0 =$ 1.0 m/s (ii) $U_0 =$ 2.0 m/s and (iii) $T_{0} \approx$ 11.5 $^{\circ}$C, $U_0 =$ 2.0 m/s. The pressure signals are centered at $t_{\text{p}} =$ 0 ms when the impact pressure at the disk center $P_{\text{c}}$ is maximum. The maximum pressure at the disk center $P_{\text{c,max}}$ at $T_{0} \approx$ 11.5 $^{\circ}$C is about 4 times larger than $T_{0} \approx$ 16.5 $^{\circ}$C, at the same impact velocity $U_0 =$ 2.0 m/s. The blue dashed lines represent the theoretical gas pressure under the approaching disk center computed with Eq. (5.4) from \cite{peters2013splash}, for comparison. (b) Pressure in the liquid bulk measured by a hydrophone located approximately 0.12 m from the center of the disk at two different ambient temperatures and impact velocities. }
\end{figure*}

The behaviour of the entrapped vapor pocket described in the previous sections directly affects the pressure exerted on the impacting disk, as well as the pressure in liquid bulk. Fig. \ref{fig:Pressure}a shows the time series of the impact pressure $p(t)$ measured at the center and near the edge of the disk at two selected $T_{0}$ and $U_0$. Here, $t_{\text{p}} =$ 0 is the time at which the peak pressure $P_{\text{c,max}}$ is recorded at the disk center. This happens shortly after the first contact between the disc and the liquid, at $t = 0$. Before impact, the pressure starts to rise because of the approaching disk pressurizing the vapor between the disk and the liquid surface (which is where the blue dashed lines and red lines overlap). In general, upon impact at $t = 0$ (i.e., $t_p < 0$), the pressure near the disk edge rises first due to the fact that the initial contact point between the disk and the deformed liquid surface lies under the disk edge. The central pressure subsequently starts to grow slightly later until it reaches a maximum. For $T_{0} \approx$ 16.5 $^{\circ}$C and $U_0 =$ 1.0 m/s in Fig. \ref{fig:Pressure}a(i), the central pressure reaches its maximum at about 1 ms after the initial pressure rise. From Fig. \ref{fig:Reff}c, we see that, for these parameter values (yellow triangular markers), the effective radius of the vapor pocket decreases rather quickly during the first 1 ms and tends toward saturation after that. Therefore, one must conclude that the maximum impact pressure at lower impact velocities occurs during the contraction of the vapor pocket. Meanwhile, for the other two depicted cases where the vapor pocket {violently} collapses, namely $T_{0} \approx$ 11.5 and 16.5 $^{\circ}$C and $U_0 =$ 2.0 m/s, the pressure build-up time is around 0.5 ms. This coincides with the time taken for the vapor pocket to retract completely, as shown by the yellow and dark blue five-pointed stars in Fig. \ref{fig:Reff}c. Thus, in these cases, the maximum pressure reached during high-velocity impact likely originated from the complete collapse of the vapor pocket. Note that these are precisely the cases where we found that the maximum central pressure at different ambient temperatures $P_{\text{c,max}}$ strongly deviates from the classic $\rho_{\text{L}}U_0^2$ scaling due to condensation of the vapor pocket that causes it to collapse violently and inducing high local impact pressure at the center of the disk, as detailed in \cite{li2025impactpressureenhancementdisk}. 

At this point, it is instructive to evaluate to what extent the measured pressure buildup could be explained from the pressurization of the vapor phase due to the approaching disk. This problem of the effect of gas cushioning on the pressure under the disk center was studied theoretically in \cite{peters2013splash}, and here, we plot the central pressure for the different conditions computed using Eq. (5.4) in \cite{peters2013splash} as the blue dashed lines in Fig. \ref{fig:Pressure}a. Clearly, the central pressure exceeds the predicted pressure with gas-cushioning effect at high impact velocity (Fig. \ref{fig:Pressure}a(ii) and (iii)), further highlighting the important role of vapor pocket condensation and collapse on the impact pressure. In addition, it should be stressed that the blue dashed lines constitute an upper limit, since the calculation is based on an inviscid vapor layer of homogeneous thickness, disregarding the effect of the Kelvin-Helmholtz instability at the disk edge.

Although the maximum impact pressure induced during a boiling liquid impact at large impact velocity and low ambient temperature is significantly higher than the impact on water in ambient air due to the condensation-induced collapse of the vapor pocket \cite{li2025impactpressureenhancementdisk}, we observe a lower impact pressure for boiling liquid impact when the vapor pocket does not collapse violently i.e., there where the $\rho_\text{L}U_0^2$ scaling is expected to hold. For example, the averaged maximum central pressure recorded for a $R_0 =$ 40 mm disk impacting into the water in ambient air is $P_{\text{c,max}} =$ 0.98 bar at $U_0 = $ 1.0 m/s \cite{jain2021air}, while under boiling liquid conditions at $T_{0} \approx$ 16.5 $^{\circ}$C and $U_0 =$ 1.0 m/s we find $P_{\text{c,max}} =$ 0.36 bar. Very likely, the thickness of the entrapped gas pocket is the factor responsible for the difference. As discussed in Section \ref{sec:Entrapment}, the thickness of the vapor pocket $h_0$ is a function of the density ratio between the gas and liquid. The density ratio for air and water is about 1.2 $\times 10^{-3}$ which is lower than the density ratio for Novec 7000 (2.3 -- 4.0 $\times 10^{-3}$) as plotted in Fig. \ref{fig:h0}b and, therefore, the thickness of the entrapped air pocket is smaller at around 115 $\mu$m \cite{jain2021total}. For water and air, the air pocket is seen to be punctured at the disk center upon impact due to high stagnation pressure and then expelled from below the disk \cite{jain2021air}. However, for Novec 7000, as shown in Fig. \ref{fig:Reff}c (yellow triangular markers), at the same impact velocity the vapor pocket retracts and remains at a nearly constant radius throughout the duration of the first central pressure peak signal ($\approx 2$ ms, from $t_{\text{p}} = -1$ to 1 ms) in Fig. \ref{fig:Pressure}a(i), providing a stronger cushioning effect and resulting in a lower impact pressure on the disk.

In addition to the disk impact pressures, the pressure in the liquid bulk $P_{\text{bulk}}$ is measured with a hydrophone placed 0.12 m away from the center of the disk. Similar to the impact pressure, as the disk impacts into the liquid, $P_{\text{bulk}}$ increases gradually until a maximum is reached. The maximum pressure $P_{\text{bulk,max}}$ is observed to increase with impact velocity, and, for the same impact velocity ($U_0 = 2.0$ m/s) is higher at $T_{0} \approx$ 11.5$^{\circ}$C as compared to $T_{0} \approx$ 16.5$^{\circ}$C. It is also seen that $P_{\text{bulk,max}}$ occurs slightly later in time after the maximum pressure at the disk center is reached, i.e., at $t_{\text{p}} >  0$. At high impact velocity ($U_0 = 2.0$ m/s), the bulk pressure signal contains multiple secondary peaks, some as large as the primary peak, a feature that is not observed at lower impact velocity. These multiple pressure peaks after impact at high impact velocity are likely caused by the propagation and reflection of the pressure waves emitted from the violent collapse of the vapor pocket and the subsequent rupture of the vapor bubble fragments. 

\subsubsection{\label{sec:Impulses} Time evolution of the impulse during boiling liquid impact}

\begin{figure*}
\includegraphics[trim=0cm 1cm 0cm 0cm, width=.7\textwidth]{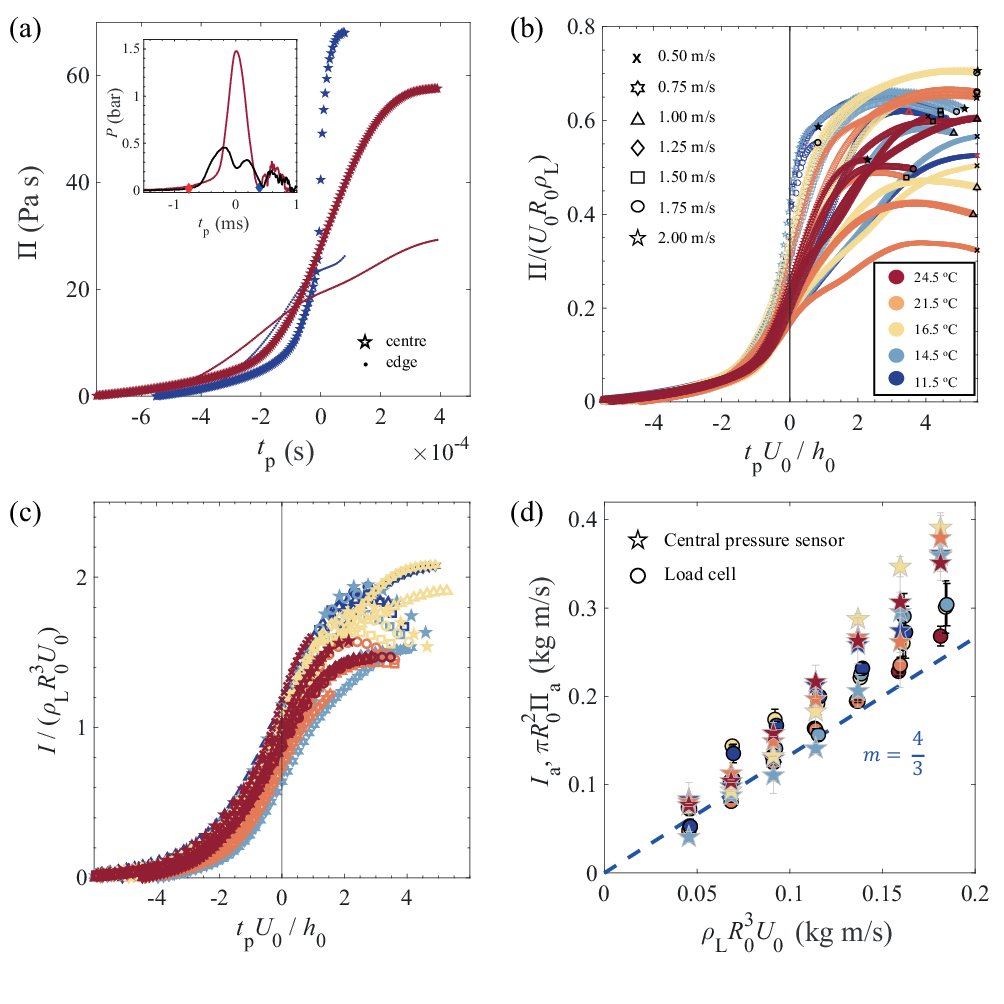}
\caption{\label{fig:Impulse} (a) Time evolution of the pressure impulses $\Pi(t)$ at the disk center (star symbols) and the disk edge (dots) for two different ambient temperatures $T_{0}\approx$ 11.5 $^{\circ}$C (blue) and 24.5 $^{\circ}$C (red) at $U_0 =$ 2.0 m/s. The inset shows the time evolution of the pressure $p(t)$ together with the typical interval of integration from $t_1$ to $t_2$ for $T_{0}\approx$ 24.5 $^{\circ}$C, $U_0 =$ 2.0 m/s. (b) Non-dimensionalized pressure impulses $\Pi/(U_0R_0\rho_{\text{L}})$ measured at the disk center for all ambient temperatures and impact velocities in the experiment plotted against the non-dimensionalized time $t_\text{p}U_0/h_0$. (c) Non-dimensionalized force impulses $I/(U_0R_0^3\rho_{\text{L}})$ versus non-dimensionalized time $t_\text{p}U_0/h_0$ for all ambient temperatures and impact velocities.  (d) Plot of the force impulses $I_\text{a}$ accumulated during the first impact peak against the momentum change associated with the added mass of liquid. The colors correspond to different $T_{0}$ as indicated in the legend in (b) but the symbols now have a different meaning. They are used to distinguish directly measured results from the (force) impulse (circles) $I_\text{a}$ from those inferred from the central pressure impulses ($\Pi_\text{a}$) by multiplying with the disk area $\pi R_0^2$. The force impulses lie closely to the theoretical calculation from \cite{batchelor2000introduction} for the force impulse transferred from an impulsively accelerated disk to an infinite half-space filled with inviscid fluid, without a gas-cushioning layer (blue dashed line). Note that the color and symbol legends in (b) equally apply to (a) and (c). }
\end{figure*}

Whereas the time evolution of the pressure gives insight into the maximum pressures occurring during impact, it is less informative if one is interested in the transfer of momentum onto the structure, in which case it is better to look at the time integral of the pressure, the pressure impulse $\Pi(t)$, together with the force impulse $I(t)$, the time integral of the force. By numerically integrating the pressure signal $p(t)$, an example of which is shown in the inset of Fig. \ref{fig:Impulse}a, from the time at which the pressure signal at the disk edge (black solid line) starts to rise $t_1$ (red diamond symbol) to the time where the first valley of the central pressure signal (red solid line) is reached $t_2$ (blue diamond symbol), the pressure impulses $\Pi(t)$ can be obtained as $\Pi(t) = \int_{t_1}^{t} p(t')\ dt'$. The time evolution of the computed pressure impulses at the disk center and edge at two different ambient temperatures $T_{0}\approx$ 11.5 $^{\circ}$C (blue) and 24.5 $^{\circ}$C (red) for $U_0 =$ 2.0 m/s are plotted in Fig. \ref{fig:Impulse}a. The pressure impulse at the disk edge starts to rise before that of the disk center, and this trend is observed for all the other impact velocities and ambient temperatures. This is distinctly different from what was observed previously during the disk impact on water in ambient air, where the pressure impulse at the disk center grows first due to a stagnation pressure build-up under the disk center \cite{jain2021air}. For boiling liquid impact, the pressure impulse at the disk edge increases steadily as the disk impacts onto the deformed liquid surface. Meanwhile, the pressure impulse at the disk center rises more slowly but overtakes that at the disk edge when getting close to $t_{\text{p}} =$ 0 due to pressurization and (in some cases) subsequent collapse of the vapor pocket. While the increase after the takeover is gradual at high ambient temperature, the growth at low ambient temperature differs discernibly as it suddenly rises sharply.

The growth of the pressure impulse before the peak pressure depends on both the liquid inertia and the intervening vapor pocket. Inertially, the pressure scales as $\rho_{\text{L}}U_0^2$ whereas the inertial time is $R_0/U_0$. The pressure impulse is therefore expected to scale as the product of the two, namely as $\rho_{\text{L}}U_0R_0$. Meanwhile, the {initial} thickness $h_0$ of the vapor pocket will dictate how long it takes to pressurize and, hence, we define a pressurization time scale as $t_{\text{c}} = h_0/U_0$.  Normalizing the pressure impulse at the disk center $\Pi$ and the time $t_{\text{p}}$ with these inertial impulse and pressurization time scales, we see that the experimental data for all the impact velocities and ambient temperatures now collapse reasonably well, especially for low $U_0$, high $T_{0}$ and during the initial loading stage before $t_{\text{p}}U_0/h_0 =$ -2. Since the same was observed for disk impact experiments on water in air \cite{jain2021air}, this suggests that the role of the vapor pocket for these early times is similar to a mediating non-condensable air pocket in this early stage. The pressure impulses arise due to the compression of the vapor pocket which naturally depends on the vapor pocket thickness $h_0$. After $t_{\text{p}}U_0/h_0 =$ -2, the pressure impulse shows clear deviation for high impact velocities ($U_0 =$ 1.75 and 2.0 m/s, circle and star symbols) and lower ambient temperature ($T_{0}\approx$ 11.5 and 14.5 $^{\circ}$C, blue shades). The deviation is likely to originate from the onset of the rapid collapse of the vapor pocket due to condensation which imparts additional instantaneous momentum on top of the impact pressurization of the vapor pocket itself. 

Now, we turn to the force measurements from the load cell that is embedded in the rod connecting the disk and the linear motor, below the seal. Note that, unlike the pressure sensor that measures local impact pressure on the disk, the force measurement is averaged over the disk and part of the rod assembly. The main features of the force signal upon impact are similar for all $U_0$ and $T_{0}$, where a first peak arises upon impact and subsequent oscillation that is most likely present inside due to the structure. The force signal time series $F(t)$ at various $U_0$ and $T_{0}$ can be found in the Supplementary Material \cite{supplemental}. What is noteworthy is that the peak impact force increases more with $U_0$ when the ambient temperature is low. The force impulses are computed in the same manner as the pressure impulse, namely by numerically integrating the force signal: $I(t) = \int_{t_1}^{t} F(t')\ dt'$. Using the same inertial pressure, length, and time scales for non-dimensionalizing the force impulse, yielding $I \sim \rho_{\text{L}}R_0^3U_0$, the force impulses are non-dimensionalized and plotted in Fig.\ref{fig:Impulse}c against the non-dimensional time $t_{\text{p}}U_0/h_0$, where again the pressurization time scale $h_0/U_0$ is taken. Again, the growth of the force impulses shows a decent collapse with the re-scaling. This further confirms that the impact load has an inertial origin and the vapor pocket thickness, which is different for each $T_{0}$ (cf. Fig. \ref{fig:h0}b) is of influence. For both the pressure and force impulses, we alternatively tried to re-scale the time with the inertial time $t_{\text{i}} = R_0/U_0$ (see the Supplementary Material \cite{supplemental}), but the data collapse is not as good as with $t_{\text{c}} = h_0/U_0$, since it creates, especially for the pressure impulse, more dispersion with the ambient temperatures. This emphasizes the significant influence of the thickness of the entrapped vapor pocket on the loading process.

Upon impact, the liquid is accelerated greatly in a short duration, which establishes a large pressure gradient that induces a sudden change in the velocity of the liquid. The magnitude of the velocity is small and hence, the approximate, linearized form of Euler's equations can be used to describe the sudden change \cite{cooker1995pressure}
\begin{equation}
\frac{\partial\textbf{u}}{\partial\text{t}} = - \frac{1}{\rho_{\text{L}}} \nabla p
\label{eq:pressure impulse}
\end{equation}
where $\textbf{u}$ is the liquid velocity. For a circular flat disk, without considering the gas-cushioning effect, this problem has been solved in \cite{batchelor2000introduction} and the resulting force impulse is $I_{\text{a}} \approx m_{\text{a}}\rho_{\text{L}}R_0^3U_0$, where the added mass coefficient equals $m_{\text{a}} = $ 4/3, and $I_a = \int^{t_2}_{t_1} F(t)\ dt$. The experimental impulses $I_{\text{a}}$ and $\Pi_{\text{a}}$ accumulated during the first impact peak are computed by taking the signals up until the local minimum closest to 0 after the force or pressure peak, i.e., $\Pi_{\text{a}} = \int^{t_2}_{t_1} p(t)\ dt$, similarly to $I_\text{a}$. The pressure impulse $\Pi_{\text{a}}$ is multiplied by the disk surface area $\pi R_0^2$ to turn it into a (force) impulse, which would correspond to the impulse exerted on the disk if it was entirely subjected to the impact pressure measured at the center. The accumulated impact force impulses from the load cell (circles) and calculated from the central pressure (stars) are plotted against the momentum change associated with the liquid added mass in Fig. \ref{fig:Impulse}d and, as expected, at low impact velocities and high ambient temperatures, the impact impulses follow the theoretical impulses $I_{\text{a}} \sim \frac{4}{3}\rho_{\text{L}}R_0^3U_0$ indicated by the blue dashed line in the plot closely. The impulses become larger than the theoretical prediction as the impact velocity increases, especially for those at low ambient temperatures and especially for those derived from the central pressure impulses. Again, this can be attributed to the rapid collapse of the vapor pocket that increases the impact impulses. Most remarkably, also for the (force) impulse, we find increased values. This is quite unexpected since the force impulse should ultimately be equal to the liquid added mass times the impact velocity, which should not depend on the presence of the phase change. Nevertheless, the force impulse is slightly larger with phase change, possibly indicating a somewhat larger transient liquid added mass due to phase change.

\section{\label{sec:Evaporation} Adding evaporation: heated disk impact}
\subsection{\label{sec:heated disk} Experimental results}

\begin{figure*}
\includegraphics[trim=0cm 0cm 0cm 0cm, width=.8\textwidth]{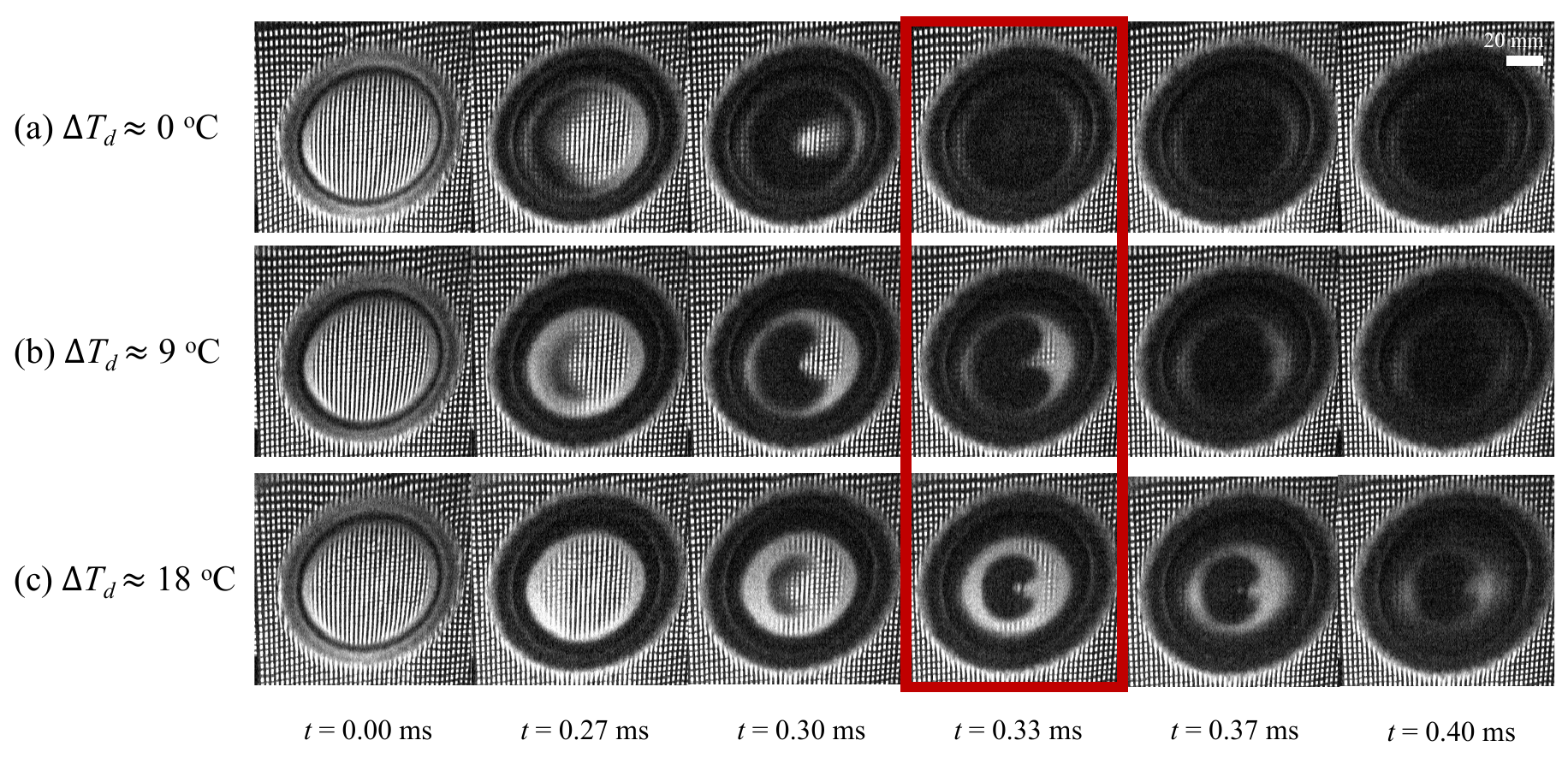}
\caption{\label{fig:bottom view heated disk} Six (re-aspected) snapshots of the bottom TIR view of the evolution of the entrapped vapor pocket at three different disk temperatures, which the disk superheat defined as $\Delta T_{\text{d}} = T_{\text{d}} - T_{0}$, where $T_{\text{d}}$ is the temperature of the heated disk and the ambient temperature $T_{0}$ is $\approx 11.5^{\circ}$C for all these cases, and $t$ is the time after impact. As highlighted by the red box, the wetted area at the central disk region reduces with increasing $\Delta T_{\text{d}}$, suggesting the retardation of the vapor pocket collapse. }
\end{figure*}

As discussed in Section \ref{sec:Dynamics} and in \cite{li2025impactpressureenhancementdisk}, we can conclude that the vapor pocket can collapse rapidly upon impact due to condensation which in turn can result in large local pressures on the impacting disk and in the liquid bulk. Trying to counteract the effect of condensation, we increase the temperature of the disk $T_{\text{d}}$  above the ambient temperature $T_0$ to see how it alters both the dynamics of the entrapped vapor pocket and the impact pressure. Here, the ambient temperature and impact velocity are fixed at $T_{0} \approx 11.5^{\circ}$C and $U_0$ = 2.0 m/s, respectively, since we observe the strongest effect of condensation at these particular parameter settings. The temperature of the disk $T_{\text{d}}$ is increased at $\approx 5^{\circ}$C intervals up to $T_{\text{d}} \approx 36^{\circ}$C, which corresponds to a maximum temperature difference with the ambient temperature $\Delta T_{\text{d}} \approx 25^{\circ}$C. Note that, for the impact experiments in this Section, as seen in Fig. \ref{fig:bottom view heated disk} (at $t = 0.00$ ms), the disk is more well-aligned compared to earlier Sections, with a tilting angle of $\alpha \leq 0.10 \pm 0.05^{\circ}$. 
%$\alpha \leq 0.0017 \pm 0.05^{\circ}$. 

\begin{figure*}
\includegraphics[trim=0cm 0cm 0cm 0cm, width=.8\textwidth]{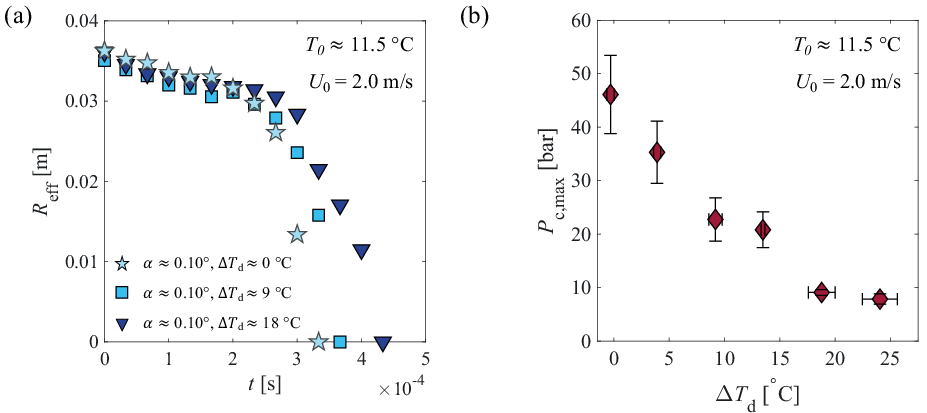}
\caption{\label{fig:heated disk} (a) Time evolution of the effective retraction radius $R_\text{eff}$ of the entrapped vapor pocket for the selected {values of the superheat $\Delta T_\text{d}$ also} shown in Fig. \ref{fig:bottom view heated disk}. Here, $\alpha$ is the tilting angle of the disk. %The black star plot is for a non-heated disk with a large tilting angle. 
(b) The maximum pressure at the disk center $P_{\text{c,max}}$ plotted against the temperature difference $\Delta T_{\text{d}}$ between the disk and the ambient.}
\end{figure*}

Referring to Fig. \ref{fig:bottom view heated disk}, by increasing the temperature of the disk, the initiation of the vapor pocket collapse is seen to be delayed. The vapor pocket starts to collapse only at $t = 0.30$ ms after the impact at $\Delta T_{\text{d}} \approx 18^{\circ}$C, which is later than the smaller $\Delta T_{\text{d}}$ cases in Fig. \ref{fig:bottom view heated disk}a and b. Besides, the expansion speed of the wetted area at the central disk region is also visibly slower than when the disk is not heated. By heating the disk, it appears that the condensation of the vapor pocket is frustrated as the collapse of the vapor pocket becomes less rapid.  The time evolution of the effective radius of the vapor pocket during heated disk impact is plotted in Fig. \ref{fig:heated disk}a. It is observed that, initially, the vapor pocket retracts at a similar rate for different $\Delta T_{\text{d}}$ up until the point the vapor pocket starts to collapse. The magnitude of the slope for $\Delta T_{\text{d}} \approx 18^{\circ}$C is smaller as it took longer for the vapor pocket to collapse completely. Most importantly, and in addition to the changes observed in the vapor pocket dynamics, also the resulting maximum impact pressure at the disk center $P_{\text{c,max}}$ is seen to decrease with the increase of $\Delta T_{\text{d}}$, as illustrated in Fig. \ref{fig:heated disk}b.

\subsection{\label{sec:evap model} Vaporization near the contact line}

We will now proceed to model the effect of heating the disk. We know that the rapid retraction of the vapor pocket in the radial direction is prompted by the condensation of the vapor at the liquid-vapor interface \cite{li2025impactpressureenhancementdisk}. Thus, the first thought that comes to mind is that conductive heat transport through the vapor layer may play a role, the magnitude of which we may estimate as $q''_\text{layer} \simeq k_\text{v}\Delta T_\text{d}/h_0$. Comparing this to the heat flux $q''_\text{L}$ into the liquid, $q_\text{L}'' \simeq k_\text{L}\Delta T_\text{v}/\sqrt{\pi\alpha_\text{L}\Delta t}$, where $\Delta t \leq h_0/U_0$, we find
\begin{equation}
    \frac{q''_\text{layer}}{q''_\text{L}} \lesssim \frac{\Delta T_\text{d}}{\Delta T_\text{v}} \frac{k_\text{v}}{k_\text{L}} \sqrt{\frac{\pi\alpha_\text{L}}{h_0U_0}} \approx 3.12 \times 10^{-3} \frac{\Delta T_\text{d}}{\Delta T_\text{v}}\ ,
\end{equation}
where we have used the relevant properties of Novec 7000 at $T_0 = 11.5^{\circ}$C ($k_\text{v} = 0.0109$ W/m K; $k_\text{L} = 0.0775$ W/m K; $\alpha_\text{L} = 4.69 \times 10^{-8}$ m$^2$/s) and $h_0 \approx 200\ \mu$m, $U_0 = 1.5$ m/s as the smaller of the two impact speeds ($U_0 = 1.5$ m/s and 2.0 m/s) used in this subsection, to arrive at the numerical value. This implies that we need to rise the temperature of the disk to $\Delta T_\text{d} \approx 321 \Delta T_\text{v}$ to obtain a heat flux through the vapor layer that is comparable to that in the liquid. Even for the modest rise of the vapor pressure that would lead to $\Delta T_\text{v} \approx 1$ K, this would imply a huge disk temperature, with $\Delta T_\text{d}$ at least an order of magnitude larger than that used in experiments. We may therefore discard heat flux through the vapor layer as the main mechanism of the heat transport from the heated disk into the liquid.

\begin{figure*}
\includegraphics[trim=0cm 0cm 0cm 0cm, width=.75\textwidth]{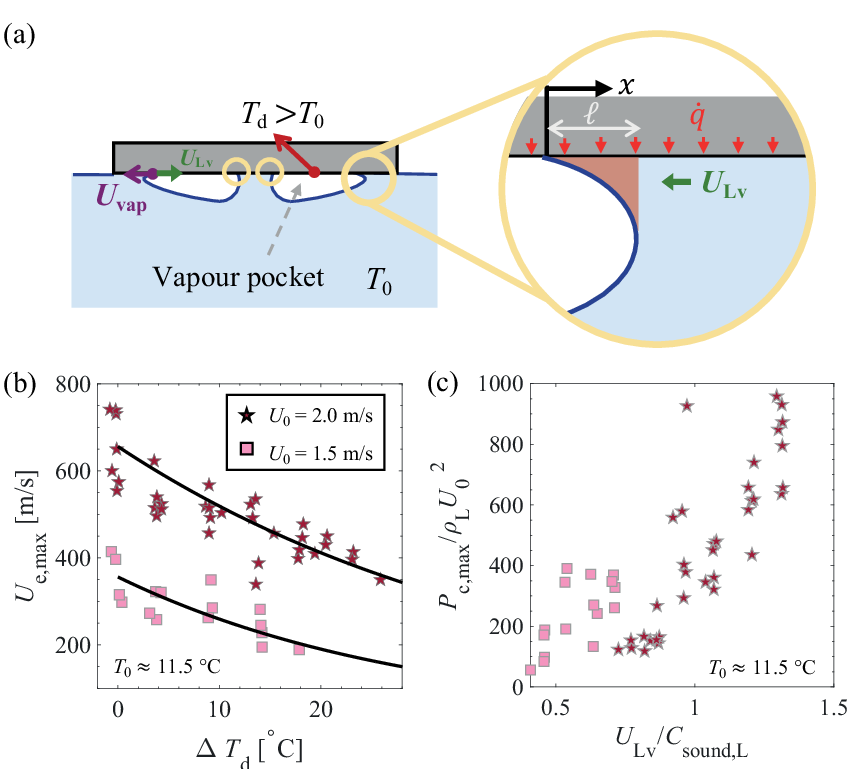}
\caption{\label{fig:model} (a) Schematic of the simplified model for the liquid-vapor movement due to the vaporization of liquid near the contact line on a heated disk. The collapse of the vapor pocket causes the liquid to rush towards the vapor at a speed $U_\text{Lv}$, while the vaporization of liquid near the three-phase contact line region causes an opposing flow at a speed $U_\text{vap}$. The red triangular shaded area indicates the contact line region, where $\ell$ is its base length. (b) Maximum expansion speed of the wetted area $U_{\text{e,max}}$ measured from the experiment at different $\Delta T_{\text{d}}$ at a fixed ambient temperature $T_0 \approx 11.5^{\circ}$C and two impact velocities ($U_0 = 1.5$ m/s and 2.0 m/s). The solid black lines represent the estimated interface speed $U_\text{Lv}$ after accounting for the effect of vaporization near the contact line based on the proposed model Eq. \eqref{eq: ULv}, using the measured average values $U_\text{Lv,0} = U_\text{e,max}(\Delta T_\text{d}=0)$ for the two impact velocities ($U_\text{Lv,0} =$ 356.1 m/s for $U_0$ = 1.5 m/s and $U_\text{Lv,0} =$ 656.0 m/s for $U_0$ = 2.0 m/s) and a single fitted value for $\ell$ (= 2.1 $\mu$m). (c) Rescaled measured pressure at the disk center $P_\text{c,max}/\rho_\text{L}U_0^2$ as a function of the fitted retraction velocity $U_\text{Lv}(\Delta T_\text{d})$, rescaled with the speed of sound $C_\text{sound,L}$ in the liquid. The legend in (b) also applies to (c).} 
\end{figure*}

Two important observations from the experiment are that the vapor pocket starts to collapse slightly later and the expansion speed of the wetted area reduces when the temperature of the disk is increased. Since the ambient temperature is fixed at $T_{0} \approx 11.5^{\circ}$C, the initial thickness of the vapor pocket $h_0$ should be the same for all experiments in this Section. Also, for all experiments with the same impact velocity $U_0$, the compression of the vapor pocket is the same, irrespective of the disk heating. On the other hand, the delay in the start time of the collapse implies a possible growth in the initial size of the vapor pocket due to heating. Meanwhile, a reduction in expansion speed suggests an opposing movement to the liquid-vapor interface at the three-phase contact line could be incurred due to vaporization. Thus, the dynamics at the three-phase contact line is expected to play the critical role in counteracting the rapid condensation-induced collapse of the vapor pocket.

To understand the mechanism on how vaporization resists the rapid collapse, here, we adopt a simplified one-dimensional model to describe the vapor generation due to heat input from the heated disk and its subsequent influence on the movement of the liquid-vapor interface. Referring to Fig. \ref{fig:model}a, upon contact between the liquid and the heated disk, either at the circumference of the vapor pocket or the center region of the disk upon the collapse of the vapor pocket (circled regions in Fig. \ref{fig:model}a), the interface will move towards the vapor at a velocity $U_{\text{in}}$ incurred by the collapse of the vapor pocket. In contrast, vaporization will deplete the liquid and induce a movement $U_{\text{vap}}$ towards the liquid (purple arrow), as if the liquid is dewetting the surface. As a result, the interface will move at a resultant speed $U_\text{Lv} = U_{\text{in}} - U_{\text{vap}}$. This resembles the process of bubble growth during nucleate boiling. The modelling approach for bubble growth during nucleate boiling to regularize the singularity at the three-phase contact line is divided into two {regions}: contact line (microregion) evaporation \cite{stephan1994new,ajaev2017heat,janevcek2012contact,nikolayev2010dynamics} and microlayer evaporation \cite{wayner1976interline,burevs2021modelling,fischer2015development} depending on several parameters, for example, the substrate superheat and velocity of the vapor-liquid interface. In contact line evaporation models, the local heat flux at the microregion (triangular shaded region in magnified Fig. \ref{fig:model}a) is considered to primarily contribute to the growth of the bubble \cite{nikolayev2022evaporation}. This is similar to our case, where due to the smaller heat conductivity of the vapor phase ($k_\text{v}/k_\text{L} \approx 0.14$ at $T_0 \approx 11.5^{\circ}$C), vaporization at the upper region of the liquid-vapor interface (where condensation is mainly occurring) due to disk heating is expected to be limited. Meanwhile, the liquid that is in contact with the heated disk but further from the three-phase contact line may directly conduct the heat into the liquid bulk. Therefore, adopting the concept of contact line evaporation, we will find $U_\text{vap}$ based on the amount of liquid that vaporizes in a region close to the three-phase contact line.

Whereas the liquid is in contact with the heated disk, the heat flux $q''$ transmitted from the heated disk to the liquid through conduction can be written as
\begin{equation} \label{eq:heatfluxGeneral}
    q'' = -k_{\text{L}} \frac{\Delta T_{\text{d}}}{\delta_{\text{t}}} = -k_{\text{L}} \frac{\Delta T_{\text{d}}}{\sqrt{\pi\alpha_{\text{L}}t}}
\end{equation}
where $k_{\text{L}}$ is the thermal conductivity of the liquid, $\delta_{\text{t}} = \sqrt{\pi\alpha_{\text{L}}t}$ is the thermal boundary layer that will grow with {the contact} time $t$ and $\alpha_{\text{L}}$ is the thermal diffusivity of the liquid. The thermal boundary grows with time as more liquid comes into contact with the heated surface. {Denoting} $x$ as the distance to the contact line and $U_\text{Lv}$ the speed of the contact line, the time in Eq. \eqref{eq:heatfluxGeneral} can be taken as $t = x/U_\text{Lv}$ since this is the time that the liquid has been in contact with the heated disk. This leads to the local heat flux $q''(x)$
\begin{equation} \label{eq:local flux}
    q''(x) = - \frac{k_{\text{L}}\Delta T_{\text{d}}}{\sqrt{\pi\alpha_{\text{L}}x}} \sqrt{U_{\text{Lv}}}\ .
\end{equation} 
Now, assuming that there is a region with a length of $x = \ell$ close to the three-phase contact line (as labelled in magnified Fig. \ref{fig:model}a), where all heat input is used for vaporization, i.e., the magnitude of the heat flow $\dot{q}(\ell)$ is given by
\begin{equation} \label{eq:aveheatflux}
    \dot{q}(\ell) = \mathcal{P} \int^{\ell}_{x= 0} |q''(x)| dx = \mathcal{P}  \int^\ell_0 \frac{k_{\text{L}}\Delta T_{\text{d}}}{\sqrt{\pi\alpha_{\text{L}}x}} \sqrt{U_{\text{Lv}}}\ dx = \mathcal{P}\frac{2k_\text{L}\Delta T_\text{d}}{\sqrt{\pi\alpha_\text{L}}}\sqrt{\ell U_\text{Lv}}
\end{equation}
where $\mathcal{P}$ is the perimeter of the contact line. This heat input is used to vaporize the liquid to generate a vapor mass at a rate of $\dot{m}_\text{v} = \dot{q}(\ell)/ L$, where $L$ is the latent heat of vaporization. From this, we can derive the resulting speed $U_\text{vap}$ of the liquid-vapor interface induced by the generation of the vapor mass at the contact line region as 
\begin{equation} \label{eq:Uevap}
    U_\text{vap} = \frac{\dot{m}_\text{v}}{\rho_\text{v,0}\ell\mathcal{P}} = \frac{\dot{q}(\ell)}{\rho_\text{v,0}\ell \mathcal{P} L} \ .
\end{equation}
Substituting the {magnitude of the heat flow} Eq. \eqref{eq:aveheatflux} into Eq. \eqref{eq:Uevap}, we eventually find
\begin{equation} \label{eq: Uevapfinal}
    U_\text{vap} = \frac{ 2k_{\text{L}}\Delta T_{\text{d}}}{\rho_\text{v,0}L\sqrt{\pi\alpha_{\text{L}}\ell}} \sqrt{U_{\text{Lv}}} \ .
\end{equation}
Finally, knowing the retraction speed $U_\text{Lv,0} = U_\text{Lv}(\Delta T_\text{d} = 0)$ for the unheated disk, we can find $U_\text{Lv}$ at any finite value of $\Delta T_\text{d}$ by substracting the speed $U_\text{vap}$ caused by the vaporization:
\begin{equation}
    U_\text{Lv} = U_\text{Lv,0} - U_\text{vap}\ .
\end{equation}
Inserting this expression in Eq. \eqref{eq: Uevapfinal} leads to a quadratic equation for $U_\text{vap}$, $U_\text{vap}^2 + BU_\text{vap} - BU_\text{Lv,0} = 0$, where $\sqrt{B}$ is defined as the prefactor of $\sqrt{U_\text{Lv}}$ in Eq. \eqref{eq: Uevapfinal}, the positive root of which provides $U_\text{vap}$, such that
\begin{equation} \label{eq: ULv}
    U_\text{Lv} = U_\text{Lv,0} - \frac{1}{2}B\left[ \sqrt{1 + \frac{4U_\text{Lv,0}}{B}}\ -1 \right] \ ,
\end{equation}
where it should be reminded that $B$ is a function of the disk superheat temperature $\Delta T_\text{d}$. The velocity of the interface towards the vapor $U_\text{Lv}$ due to the collapse of the vapor pocket without any heating can be obtained experimentally for an unheated disk $U_\text{e,max}(\Delta T_\text{d} = 0) $ or from the theoretical expression for $U_{\text{in}}$ derived in \cite{li2025impactpressureenhancementdisk} based on the condensation of the vapor pocket. Hence, the only unknown that needs to be found out now is $\ell$, which is the length near the three-phase contact line where vaporization actually occurs.

Fig. \ref{fig:model}b shows the measured maximum expansion speed $U_\text{e,max}$ of the wetted area around the central region for disks of different $\Delta T_{\text{d}}$ at  $T_0 \approx 11.5^{\circ}$C, quantified in the same way as described in Section \ref{sec:Expansion}. Additional $U_{\text{e,max}}$ data for the experiment at $U_0 = 1.50$ m/s and $T_0 \approx 11.5^{\circ}$C is shown in Fig. \ref{fig:model}b as well (square symbols), where the vapor pocket is also seen to collapse from the center region of the disk in the experiment, but at a slower speed. At $U_0 = 2.0$ m/s, when $\Delta T_{\text{d}} \approx 0$$^{\circ}$C, $U_{\text{e,max}}$ is about 650 m/s and decreases gradually to about 350 m/s with increasing $\Delta T_{\text{d}}$. The liquid-vapor interface speed $U_\text{Lv}$ in Eq. \eqref{eq: ULv} is equivalent to $U_\text{e,max}$. To obtain $\ell$, we fit the experimental $U_{\text{e,max}}$ to Eq. \eqref{eq: Uevapfinal} using {$U_\text{Lv,0} = U_\text{e,max,0}$, computed as the average of data points with $|\Delta T| < $ 0.3 $^{\circ}$C}. This results in a length scale in the order of $\mu$m that matches well with the data. The solid lines in Fig. \ref{fig:model}b show the fitted Eq. \eqref{eq: ULv} taking $\ell = 2.1\ \mu$m for both $U_0 = 1.5$ m/s and 2.0 m/s. It appears that this length of the contact line region is not strongly dependent on the impact velocity, at least at the same $T_0$.

Additionally, in Fig. \ref{fig:model}c, we plot the normalized maximum pressure at the disk center $P_\text{c,max}/\rho_\text{L}U_0^2$ against the normalized inward retraction speed on a heated disk $U_\text{Lv}/C_\text{sound,L}$, computed by (again) taking $U_\text{Lv,0} = U_\text{e,max,0}$ and $\ell = 2.1\ \mu$m in Eq. \eqref{eq: ULv}. Previously, we have demonstrated that the maximum impact pressure $P_\text{c,max}/\rho_\text{L}U_0^2$ at the disk center scales reasonably well with the condensation-induced inward radial retraction speed $U_\text{in}/C_\text{sound,L}$ of the vapor pocket at different ambient temperatures and impact velocities \cite{li2025impactpressureenhancementdisk}. Clearly, there is also a dependency of the maximum impact pressure on the resultant inward retraction speed $U_\text{Lv}$ of the vapor pocket during heated disk impact, although more data dispersion is present in the latter case. As the violent collapse of the vapor pocket due to rapid condensation is resisted by the vaporization of liquid near the contact line, the impact pressure also decreases in response to the reduced retraction speed. The proposed model describes the movement of the liquid-vapor interface due to heating reasonably well, supporting our argument that vaporization over length $\ell$ near the contact line region supplies additional vapor to the vapor pocket and induces a retraction of the liquid upon vaporization $U_\text{vap}$ that slows down the rapid collapse of the vapor pocket to $U_\text{Lv}$, and hence, lowering the resulting local impact pressure on the disk. 

\section{\label{sec:tilt} Effect of tilting angle of the disk}

\begin{figure*}
\includegraphics[trim=0cm 0cm 0cm 0cm, width=0.8\textwidth]{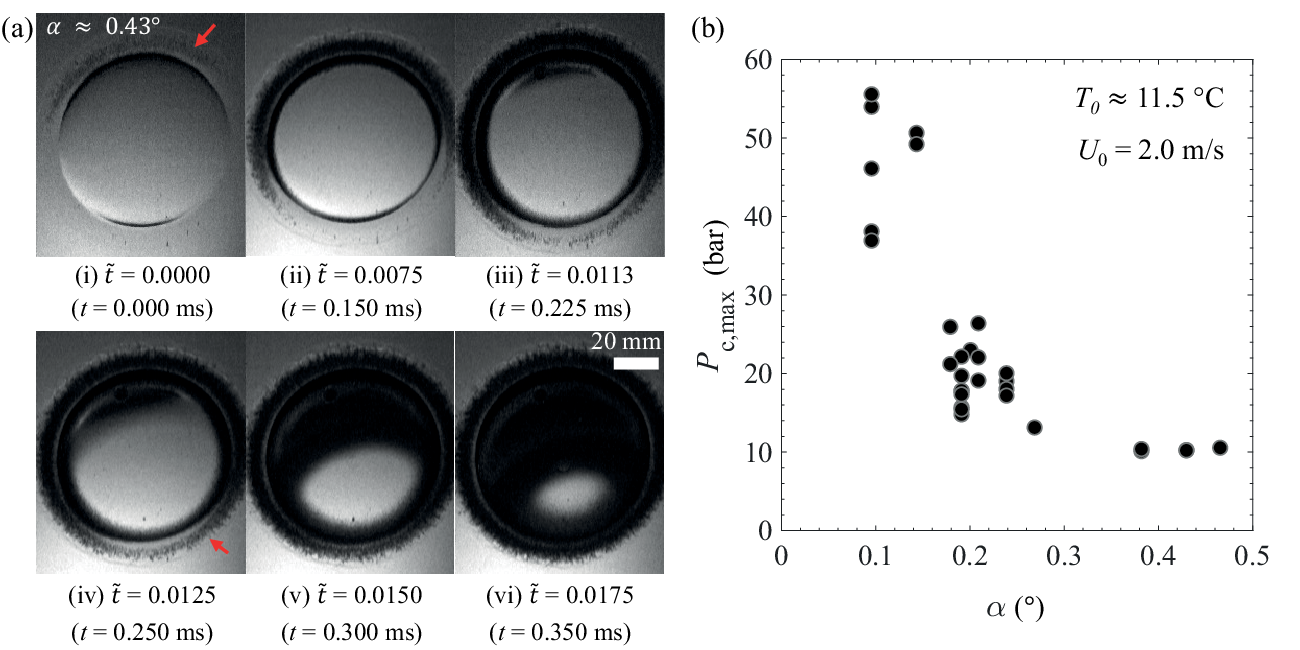}
\caption{\label{fig:tilt} (a) Six (re-aspected) snapshots of the bottom view of an impacting disk with a large tilting angle $\alpha \approx 0.43\ \pm\ 0.04 ^{\circ}$ at $U_0 = 2.0$ m/s and $T_{0} \approx 11.5 ^{\circ}$C. Two time labels are shown where $t$ is the dimensional time after impact and $\tilde{t}$ is time non-dimensionalized with the inertial time scale ($\tilde{t} = U_0t/R_0$) for the ease of comparison with Fig. \ref{fig:heated disk}a(i) ($\alpha \approx 0.10 ^{\circ}$) and Fig. \ref{fig:h0}a(i) ($\alpha \approx 0.25 ^{\circ}$). 
(b) Maximum central pressure $P_{\text{c,max}}$ for individual repetitions of the impact at $U_0 = 2.0$ m/s and $T_{0} \approx 11.5 ^{\circ}$C against tilt angle $\alpha$. }
\end{figure*}

In the experiments discussed in the previous sections, it was clear that the tilting angle plays a major role in determining the pressures recorded, especially for parameter settings in which a violent collapse of the vapor pocket is expected. For instance, for the $T_0 \approx 11.5^{\circ}$C, $\Delta T_\text{d} = 0^{\circ}$, and $U_0 = 2.0$ m/s experiments in Section \ref{sec:Evaporation}, a maximum impact pressure $P_\text{c,max} \approx 45$ bar was recorded with a tilt angle $\alpha \approx 0.10^{\circ}$, whereas in the other sections, at $\alpha \approx 0.25^{\circ}$ and the same settings, we found $P_\text{c,max} \approx 18$ bar, that is, a factor of 2.5 smaller. Therefore, in this section we will address the effect of the tilting angle $\alpha$ of the impacting disk on the dynamics of the vapor pocket and the impact pressure in detail. Here, the ambient temperature and the impact velocity are fixed at $T_{0} \approx 11.5^{\circ}$C and $U_0 = 2.0$ m/s, where the vapor pocket collapses most violently in our range of experimental settings, while the tilting angle of the disk is varied. Referring to Fig. \ref{fig:tilt}a, the tilting angle is estimated from the bottom view recordings using $\alpha = \text{arctan} (U_0\Delta t/2R_0)$, where $\Delta t $ is the time it takes from the moment of initial contact at one end of the disk edge (see red arrow in a(i), $t = 0$ ms) until the moment at which the other end of the disk (red arrow in a(iv)) is completely in contact with the liquid (at $t$ = 0.300 ms). The tilted disk initially contacts the liquid from the top edge (red arrow in (i)), which is the same for the case shown in Fig. \ref{fig:Reff}a(ii) where $\alpha \approx 0.25 \pm 0.05^{\circ}$. However, due to the larger $\alpha$ in this case ($\alpha \approx 0.43 \pm 0.04 ^{\circ}$), a wetted patch starts to move from the top region at $t = 0.15$ ms towards the bottom, before the other end is fully in contact with the liquid, and subsequently spreads rapidly towards the other end of the disk, which is a different behaviour from that showed in Section \ref{sec:Dynamics}, where the wetting take place in the central region.

Impact of a tilted cylindrical body onto water in ambient air is studied in \cite{speirs2021cavitation}, where the contact line movement from one end to the other upon impact is found to induce cavitation due to a compression wave formed with the moving contact line that compresses the liquid. Eventually, the wave reflects at the free surface and forms a tension wave, creating a negative pressure zone under the disk when cavitation occurs. Such movement of the contact line was not observed at a very low tilting angle $\alpha \approx 0^{\circ}$ in \cite{speirs2021cavitation}, and cavitation was not incurred due to the entrapped air pocket that effectively cushioned the impact. This is consistent with our own findings that the tilting angle of $\alpha \approx 0.25\ \pm\ 0.05^{\circ}$ in the experimental results shown in Section \ref{sec:Dynamics} does not influence the proper entrapment of the vapor pocket. For a tilted disk, the vapor pocket is technically not enclosed yet upon initial impact, and vapor could escape from the unclosed end. Fig. \ref{fig:tilt}b depicts the maximum impact pressure $P_\text{c,max}$  measured at the center of the disk for the disks with different tilting angles $\alpha$ at the same settings. Comparing $P_{\text{c,max}}$ at different $\alpha$, we observe that $P_{\text{c,max}}$ decreases when the tilting angle of the disk increases. The pressure in the liquid bulk measured by the hydrophone also displays the same trend (see the Supplementary Material \cite{supplemental}). It may be counterintuitive at first as at small $\alpha$, the vapor pocket is well-entrapped (and thus may cushion the impact if there is no condensation) but $P_\text{c,max}$ is higher compared to large $\alpha$ where the vapor can escape through the unclosed end. However, during boiling liquid impact, or the chosen parameter settings the effect of phase change is dominant. As discussed previously, the compression of the entrapped vapor pocket induces condensation, resulting in the rapid collapse of the vapor pocket, because the pocket is strongly pressurized ($\sim \rho_\text{L}C_\text{sound,L}U_0$) \cite{li2025impactpressureenhancementdisk}. A tilted disk may interfere with the compression buildup with an unclosed end initially at one end of the disk. As seen in Fig. \ref{fig:tilt}a, the collapse of the vapor pocket at the central region is not seen for a tilted disk and $P_\text{c,max}$ is lower than the cases where the vapor pocket collapses. Therefore, the tilt angle of the disk can influence the entrapment and subsequent compression of the vapor pocket, hindering the rapid condensation of the vapor pocket and resulting in a lower impact pressure. Meanwhile, for the range of tilt angles where the vapor pocket is properly entrapped ($\alpha < 0.30^{\circ}$), the degree of tilting has an influence on the final location of the collapse of the vapor pocket. As seen in Fig. \ref{fig:bottom view heated disk}a, when the disk is well-aligned $(\alpha \approx 0.10^{\circ})$, the final collapse of the vapor pocket is directly at the center of the disk where the pressure sensor is located, thereby giving the highest pressure of almost 50 bar. In short, above some threshold tilt angle, the movement of the three-phase contact line over the disk surface will affect the proper entrapment and, therefore, the subsequent dynamics of the vapor pocket and the resulting impact pressure on the disk. 

Here it is good to note that the tilt angle of the disk arises from the alignment of the linear motor and the main chamber of the setup, which implies that although it is not difficult to measure, it is very hard to control, and values of $\alpha \approx 0.10^{\circ}$ are extremely hard to obtain.

\section{\label{sec:Conclusion} Conclusion and Remarks}
To conclude, we have investigated the dynamics of the vapor pocket that is entrapped under a circular flat disk during impact in a boiling liquid system, i.e., a system where the liquid is in thermal equilibrium with its vapor. Through a series of experiments, we shed light on how several parameters, including the ambient temperature, the impact velocity, temperature and impact angle of the disk result in the different dynamics of the entrapped vapor pocket, which have a direct influence on the resulting hydrodynamics load. The vapor pocket in general retracts from the disk edge as the disk descends. We show that the retraction of the vapor pocket at the initial stage is driven by dynamic pressure, resembling a Rayleigh collapse. At a later stage, for impact at high-impact velocity and low ambient temperature of the system, the thinner vapor pocket is highly compressed and condenses quickly at the liquid-vapor interface, causing it to collapse rapidly and completely from the center of the vapor pocket. The collapse of the vapor pocket can be supersonic, exceeding the speed of sound in the liquid. While impact at high impact velocity and low ambient temperature results in high impact pressure on the disk due to the condensation-induced collapse of the vapor pocket, we highlight that, in other cases where the vapor pocket does not collapse violently, the impact pressure during boiling liquid impact is lower than in water-air impact. This is due to the larger density ratio between the vapor and liquid phase compared to the water-air case, that deforms the liquid surface more before impact, entrapping a thicker vapor pocket that sustains through the pressurization period $t_{\text{c}} = h_0/U_0$, leading to more effective cushioning of the impact. The pressure and force impulses during impact are found to still originate from the acceleration of the added mass, with the collapse of the vapor pocket imparting instantaneous pressure impulses that do not strongly influence the global behaviour. In addition, by heating the impacting disk, we show that the rapid collapse of the vapor pocket can be resisted by the vaporization of the liquid near the three-phase contact line that imposes an opposing motion to the spreading of the condensed liquid, which in return, slows down the collapse of the vapor pocket and thereby reduces the high impact pressure on the disk caused by the rapid condensation-induced collapse of the vapor pocket. Finally, a high tilting angle ($> 0.30^{\circ}$) of the disk is observed to obstruct the proper entrapment and the subsequent pressurization of the entrapped vapor pocket, resulting in a lower impact pressure due to the absence of condensation-induced collapse of the vapor pocket around the central disk region. 

As a closing remark, significant progress has been made by the scientific community in unravelling the complex physics of solid–liquid impact across a wide range of scales since the pioneering work of Worthington \cite{worthington1877xxviii} almost two centuries ago. The fundamental understanding — encompassing the role of liquid compressibility \cite{wagner1932stoss,korobkin2007second,korobkin2006numerical,lesser1983impact}, gas layer dynamics \cite{bagnold1939interim,bouwhuis2015initial}, solid elasticity \cite{khabakhpasheva2013fluid,todter2020experimentally}, and interfacial instabilities \cite{van2022linear,jain2021KH} — has laid the foundation for addressing more intricate solid-liquid impact phenomena. By incorporating phase change, our study extends this body of knowledge into the domain of boiling liquid impact that is of particular relevance to cryogenic fuel containment systems. It is also closely related to multiphase pipeline transmission, such as CO2 injection where water hammer phenomena are
often encountered upon abrupt change in fluid flow, often in conjunction with condensation \cite{gruel1980steam,wang2018experimental,star2025temperature}. Our experimental results reveal that the dynamics of entrapped vapor differ fundamentally from non-condensable air, owing to the propensity of vapor to undergo condensation. The condensation-induced collapse of the entrapped vapor results in exceptionally high local impact pressures that can exceed the limiting pressure from the Joukowsky equation, i.e., the water hammer pressure $\rho_\text{L}C_\text{sound,L}U_0$. This effect of condensation-induced collapse of the entrapped vapor is even more pronounced at large-scale wave impact, where the pressure can reach almost 100 bar  \cite{ezeta2025large}, i.e., two orders of magnitude larger than in the air-water case. On top of the influence of the impact scale, our studies also highlight the critical influence of the size of the vapor cavity on the resulting impact loads. Therefore, moving on, it is worth investigating how other forms of vapor, for instance, small vapor bubbles or vapor bubble clouds, can alter the overall impact loading. A better understanding of the physics of boiling liquid impact will move us a step closer to developing a physics-driven, predictive model for assessing and mitigating impact loads in next-generation cryogenic fuel containment systems, such as liquid hydrogen, extending or even replacing the trial-and-error approach in incorporating phase change that is currently in use.

\begin{acknowledgments}
The authors express their gratitude to J.M. Gordillo and A. Prosperetti for insightful discussions. We also thank G.-W. Bruggert, M. Bos and T. Zijlstra for their technical support. Special thanks to D. van Gils for setting up the disk heating system. This work is part of the Vici project IMBOL (project number 17070) which is partly financed by the Dutch Research Council (NWO).
\end{acknowledgments}

\section*{DATA AVAILABILITY}
The data that support the findings of this article are openly available \cite{dataSet}.

\bibliography{ref}

\end{document}

\title{Supplemental Material\\{\small Disk impact on a boiling liquid: Dynamics of the entrapped vapor pocket}\\}

\author{Yee Li (Ellis) Fan$^1$}
\email{Contact author: ellisfan179@gmail.com}
\author{Bernardo Palacios Muñiz$^1$}
\author{Nayoung Kim$^{1}$} 
\author{Devaraj van der Meer$^1$}
\email{Contact author: d.vandermeer@utwente.nl}

\affiliation{% 
$^1$Physics of Fluids Group and Max Planck Center Twente for Complex Fluid Dynamics,
MESA+ Institute and J. M. Burgers Centre for Fluid Dynamics, University of Twente,
P.O. Box 217, 7500AE Enschede, The Netherlands}%

% \date{\today}

%\begin{abstract}
%\end{abstract}


\maketitle

In this Supplementary Material, we provide some figures that complement the ones in the main text, either by providing additional data (Supplementary Material \ref{app:full Reff}) or by showing quantities that were mentioned in the main text but not explicitly shown (Supplementary Material \ref{app:load cell} and \ref{app:tilted hydro}). In addition, in Supplementary Material \ref{app:load cell} we also provide an alternative rescaling of the pressure and force impulse data. 

\subsection{Full $R_{\text{eff}}$ and normalized $R_{\text{eff}}$ plots} \label{app:full Reff}
\begin{figure*}[h]
\includegraphics[trim=0cm 0cm 0cm 0cm, width=.7\textwidth]{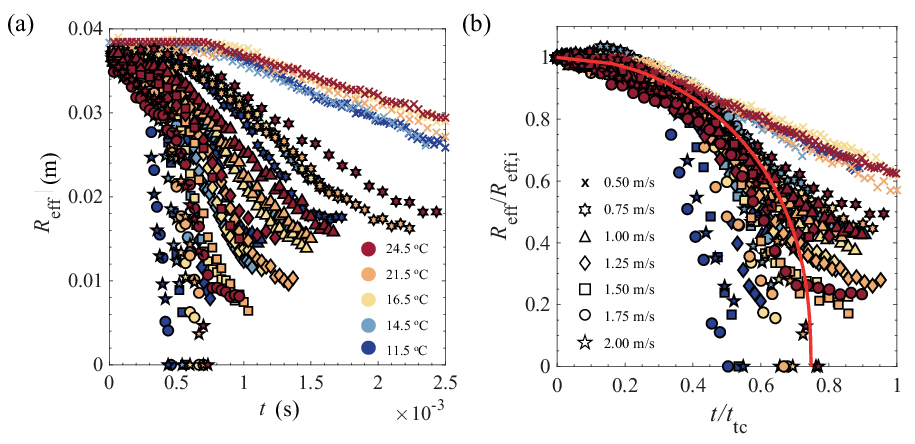}
\caption{\label{fig:full Reff} Plot of (a) $R_{\text{eff}}$ against time and (b) $R_{\text{eff}}$ and time normalized by initial effective radius $R_{\text{eff,i}}$ and the time needed for complete collapse $t_{\text{tc}}$, respectively, for all impact velocities (symbols) and ambient temperatures (colours) conducted in the experiment. The red solid line is the solution of Eq. (2) in the main text. The legends apply to both figures. }
\end{figure*}

In Section IV C of the main text, we have shown the effective radius $R_\text{eff}$ of the vapor pocket as a function of time $t$ for several representative ambient temperatures and impact velocities. Here, in Fig. \ref{fig:full Reff}a, we provide the full plot of $R_\text{eff}$ versus $t$ for all the ambient temperatures and impact velocities covered in our experiments. Meanwhile, Fig. \ref{fig:full Reff}b shows the normalized plot of Fig. \ref{fig:full Reff}a, where $R_\text{eff}$ is normalized by its initial radius $R_\text{eff,i}$ and $t$ is normalized by the Rayleigh collapse time scale $t_\text{tc}$ discussed in Section IV C in the main text.

\subsection{Load cell signal and alternative time scale for impulses}
\label{app:load cell}

\begin{figure*}[h]
\includegraphics[trim=0cm 0cm 0cm 0cm, width=1\textwidth]{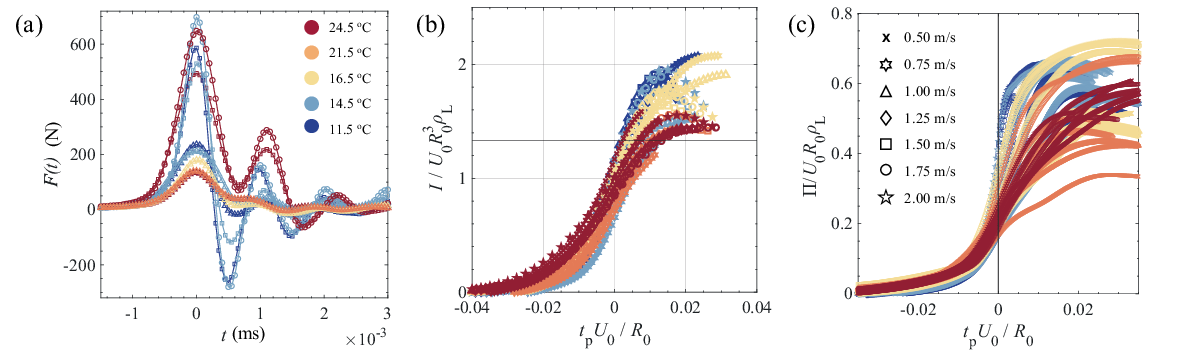}
\caption{\label{fig:load cell signal} (a) Force time series measured by the load cell for selected impact velocities and ambient temperature. Note that, no data is available for $U_0 = 2.0$ m/s at $T_{0} \approx 11.5^{\circ}$C due to the force exceeding the capacity of the load cell (890N). (b) Pressure and (c) force impulse data that is shown in Fig. 6b and 6c of the main text, normalized by the inertial time scale $t = R_0/U_0$. The collapse of the curves is not as good as compared to the compression time scale $t = h_0/U_0$. The legends are for all three figures.}
\end{figure*}

Fig. \ref{fig:load cell signal}a shows several force signal time series $F(t)$ at different impact velocities $U_0$ and ambient temperature $T_{0}$, that were used for computing the force impulses $I$ in Section IV D 2 of the main text. In general, the trend of impact force is the same as for the impact pressure, where the impact force increases with reducing ambient temperature and increasing impact velocity. What is different is that after attaining the maximum, there is a clear oscillation visible in $F(t)$, which is absent in the pressure signals and is observed to have a frequency that is mostly independent of impact parameters. We therefore attribute this oscillation to the disk-rod system, rather than being connected to the impact or vapor pocket dynamics.

In the main text, we adopted the compression time scale $t = h_0/U_0$ as the appropriate time scale for the pressure and force impulses. Here, we normalized the pressure and force impulse data with the inertial time scale $t = R_0/U_0$, as shown in Fig. \ref{fig:load cell signal}b and c. The highest (red) and lowest (blue) ambient temperature data do not collapse as good with this inertial time scale due to the difference in the initial vapor pocket thickness $h_0$, which makes the compression time scale more suitable to represent the data.

\subsection{Hydrophone pressure measurement for tilted disk impact}
\label{app:tilted hydro}

\begin{figure*}[h]
 \includegraphics[trim=0cm 0cm 0cm 1cm, width=.4\textwidth]{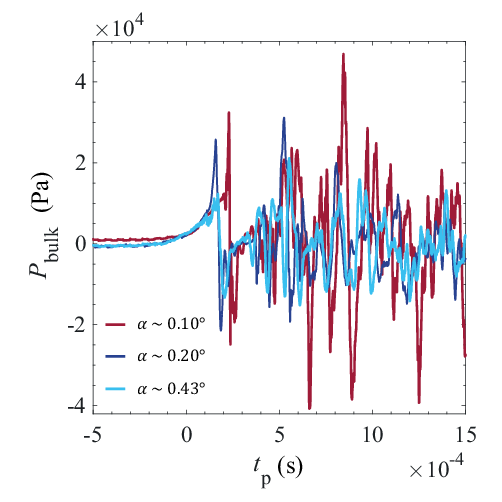}
\caption{\label{fig:tilted hydro} Hydrophone pressure measurement for the three selected cases of tilted disk impact. }
\end{figure*}

Fig. \ref{fig:tilted hydro} shows the hydrophone measurement for three selected tilt angles $\alpha \approx 0.10^{\circ}, 0.25^{\circ}\ \text{and} \ 0.43^{\circ}$ that was discussed in Section  VI of the main text. Similar to the maximum pressure at the disk centre, the initial pressure rise in the liquid bulk increases with decreasing tilting angle of the disk $\alpha$.